# Universal Characteristics of Fractal Fluctuations in Prime Number Distribution

A. M. Selvam

Deputy Director (Retired)
Indian Institute of Tropical Meteorology, Pune 411 008, India
Web sites: http://www.geocities.com/amselvam
http://amselvam.tripod.com/index.html
Email: amselvam@gmail.com

The frequency of occurrence of prime numbers at unit number spacing intervals exhibits selfsimilar fractal fluctuations concomitant with inverse power law form for power spectrum generic to dynamical systems in nature such as fluid flows, stock market fluctuations, population dynamics, etc. The physics of long-range correlations exhibited by fractals is not yet identified. A recently developed general systems theory visualises the eddy continuum underlying fractals to result from the growth of large eddies as the integrated mean of enclosed small scale eddies, thereby generating a hierarchy of eddy circulations, or an inter-connected network with associated long-range correlations. The model predictions are as follows: (i) The probability distribution and power spectrum of fractals follow the same inverse power law which is a function of the golden mean. The predicted inverse power law distribution is very close to the statistical normal distribution for fluctuations within two standard deviations from the mean of the distribution. (ii) Fractals signify quantumlike chaos since variance spectrum represents probability density distribution, a characteristic of quantum systems such as electron or photon. (ii) Fractal fluctuations of frequency distribution of prime numbers signify spontaneous organisation of underlying continuum number field into the ordered pattern of the quasiperiodic Penrose tiling pattern. The model predictions are in agreement with the probability distributions and power spectra for different sets of frequency of occurrence of prime numbers at unit number interval for successive 1000 numbers. Prime numbers in the first 10 million numbers were used for the study.

## 1. Introduction

Dynamical systems in nature such as atmospheric flows, heartbeat patterns, population dynamics, stock market indices, DNA base A, C, G, T sequence pattern, prime number distribution, etc., exhibit irregular (chaotic) space-time fluctuations on all scales and exact quantification of the fluctuation pattern for predictability purposes has not yet been achieved. The irregular fluctuations, however manifest a new kind of order, that of selfsimilarity, i.e., the larger scale fluctuations resemble in shape the enclosed smaller scale fluctuations signifying long-range correlations seen as inverse power law form for power spectra of the fluctuations. The fractal or selfsimilar nature of space-time fluctuations was identified by Mandelbrot (1975) in the 1970s. Representative examples of fractal fluctuations of prime number distribution are shown Fig. 1. Power-law behavior in the distribution of primes and correlations in prime numbers have been found (Wolf, 1997), along with multifractal features in the distances between consecutive primes (Wolf, 1989). Many physical and biological systems exhibit patterns where prime numbers play an important role (Ares and Castro, 2006). Examples range from the periodic orbits of a system in quantum chaos to the life cycles of species (Kumar *et al*, 2008). The 'number theory and physics archive' maintained by Watkins (2008) makes available an ever-expanding web-resource documenting the curious, emerging interface between these subjects.

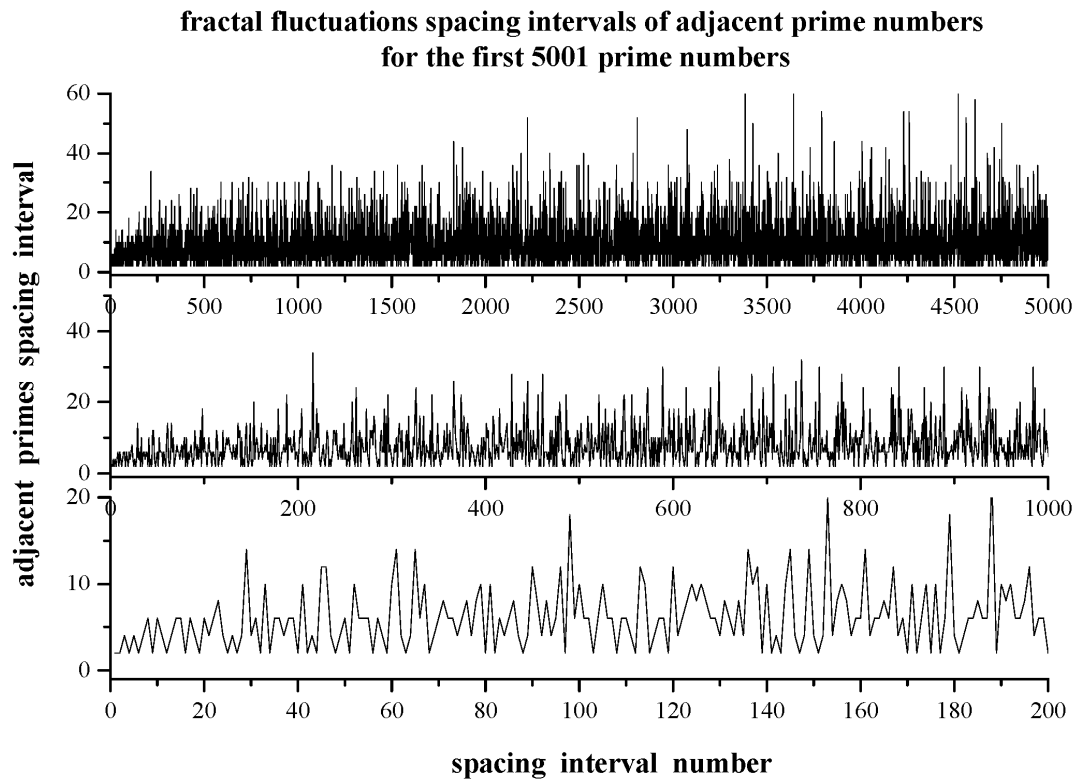

**fractal fluctuations spacing intervals of adjacent prime numbers
for the first 5001 prime numbers**

Fig. 1: The zigzag selfsimilar pattern of fractal fluctuations exhibited by the spacing intervals of adjacent prime numbers at different resolutions

Fractal fluctuations (Fig. 1) show a zigzag selfsimilar pattern of successive increase followed by decrease on all scales (space-time), for example in atmospheric flows, cycles of increase and decrease in meteorological parameters such as wind, temperature, etc. occur from the turbulence scale of millimeters-seconds to climate scales of thousands of kilometers-years. The power spectra of fractal fluctuations exhibit $1/f$ noise (Planat, 2003) manifested as inverse power law of the form $f^\alpha$ where $f$ is the frequency and $\alpha$ is a constant. An extensive bibliography of publications on $1/f$ noise in complex systems is given by Li (2008). Inverse power law for power spectra indicate long-range space-time correlations or scale invariance for the scale range for which $\alpha$ is a constant, i.e., the amplitudes of the eddy fluctuations in this scale range are a function of the scale factor $\alpha$ alone. In general the value of $\alpha$ is different for different scale ranges indicating multifractal structure for the fluctuations. The long-range space-time correlations exhibited by dynamical systems are identified as self-organized criticality (Bak *et al*., 1988; Schroeder, 1990). $1/f$ fluctuations occur in areas as diverse as electronics, chemistry, biology, cognition or geology and claims for an unifying mathematical principle (Milotti, 2002; Planat *et al*., 2002; Planat, 2003). The physics of fractal fluctuations generic to dynamical systems in nature is not yet identified and traditional statistical, mathematical theories do not provide adequate tools for identification and quantitative description of the observed universal properties of fractal structures observed in all fields of science and other areas of human interest. A recently developed general systems theory for fractal space-time fluctuations (Selvam, 1990, 2005, 2007; Selvam and Fadnavis, 1998) shows that the larger scale fluctuation can be visualized to emerge from the space-time averaging of enclosed small scale fluctuations, thereby generating a hierarchy of selfsimilar fluctuations manifested as the observed eddy continuum in power spectral analyses of fractal fluctuations. Such a concept results in inverse power law form incorporating the *golden mean* τ for the space-time fluctuation pattern and also for the power spectra of the fluctuations (Sec. 4). The predicted distribution is close to the Gaussian distribution for small-scale fluctuations, but exhibits *fat long tail* for large-scale fluctuations. The general systems theory, originally developed for turbulent fluid flows, provides universal quantification of physics underlying fractal fluctuations and is applicable to all dynamical systems in nature independent of its physical, chemical, electrical, or any other intrinsic characteristic. In the following, Sec. 2 gives a summary of traditional statistical and mathematical theories/techniques used for analysis and quantification of space-time fluctuation data sets. The drawbacks of existing techniques of data quantification are discussed and important model predictions of the general systems theory are listed. The applications of general systems theory concepts to number theory in general and to prime number distribution in particular are discussed in Sec.3. The general systems theory for fractal space-time fluctuations originally developed for turbulent fluid flows is described in Sec. 4 and the universal Feigenbaum's constants *a* and *d* characterizing dynamical systems is incorporated in model predictions in Sec.5. Sec. 6 deals with data and analyses techniques. Discussion and conclusions of results are presented in Sec. 7.

## 2. Statistical methods for data analysis

Dynamical systems such as atmospheric flows, stock markets, heartbeat patterns, population growth, traffic flows, etc., exhibit irregular space-time fluctuation patterns. Quantification of the space-time fluctuation pattern will help predictability studies, in particular for events which affect day-to-day human life such as extreme weather events, stock market crashes, traffic jams, etc. The analysis of data sets and broad quantification in terms of probabilities belongs to the field of statistics. Early attempts resulted in identification of the following two quantitative (mathematical) distributions which approximately fit data sets from a wide range of scientific and other disciplines of study. The first is the well known statistical normal distribution and the second is the power law distribution associated with the recently identified '*fractals*' or selfsimilar characteristic of data sets in general. Traditionally, the Gaussian probability distribution is used for a broad quantification of the data set variability in terms of the sample mean and variance. In the following, a summary is given of the history and merits of the two distributions.

### 2.1 Statistical normal distribution

Historically, our present day methods of handling experimental data have their roots about four hundred years ago. At that time scientists began to calculate the odds in gambling games. From those studies emerged the theory of probability and subsequently the theory of statistics. These new statistical ideas suggested a different and more powerful experimental approach. The basic idea was that in some experiments random errors would make the value measured a bit higher and in other experiments random errors would make the value measured a bit lower. Combining these values by computing the average of the different experimental results would make the errors cancel and the average would be closer to the "right" value than the result of any one experiment (Liebovitch and Scheurle, 2000).

  Abraham de Moivre, an 18th century statistician and consultant to gamblers made the first recorded discovery of the normal curve of error (or the bell curve because of its shape) in 1733. The normal distribution is the limiting case of the binomial distribution resulting from random operations such as flipping coins or rolling dice. Serious interest in the distribution of errors on the part of mathematicians such as Laplace and Gauss awaited the early nineteenth century when astronomers found the bell curve to be a useful tool to take into consideration the errors they made in their observations of the orbits of the planets (Goertzel and Fashing, 1981, 1986). The importance of the normal curve stems primarily from the fact that the distributions of many natural phenomena are at least approximately normally distributed. This normal distribution concept has molded how we analyze experimental data over the last two hundred years. We have come to think of data as having values most of which are near an average value, with a few values that are smaller, and a few that are larger. The probability density function PDF($x$) is the probability that any measurement has a value between $x$ and $x$ + d$x$. We suppose that the PDF of the data has a normal distribution. Most quantitative research involves the use of statistical methods presuming *independence* among data points and Gaussian 'normal' distributions (Andriani and McKelvey, 2007). The Gaussian distribution is reliably characterized by its stable mean and finite variance (Greene, 2002). Normal distributions place a trivial amount of probability far from the mean and hence the mean is representative of most observations. Even the largest deviations, which are exceptionally rare, are still only about a factor of two from the mean in either direction and are well characterized by quoting a simple standard deviation (Clauset, Shalizi, and Newman, 2007). However, apparently rare real life catastrophic events such as major earth quakes, stock market crashes, heavy rainfall events, etc., occur more frequently than indicated by the normal curve, i.e., they exhibit a probability distribution with a *fat tail*. Fat tails indicate a power law pattern and interdependence. The "tails" of a power-law curve — the regions to either side that correspond to large fluctuations — fall off very slowly in comparison with those of the bell curve (Buchanan, 2004). The normal distribution is therefore an inadequate model for extreme departures from the mean.

The following references are cited by Goertzel and Fashing (1981, 1986) to show that the bell curve is an empirical model without supporting theoretical basis: (i) Modern texts usually recognize that there is no theoretical justification for the use of the normal curve, but justify using it as a convenience (Cronbach, 1970). (ii) The bell curve came to be generally accepted, as M. Lippmnan remarked to Poincare (Bradley, 1969), because "...the experimenters fancy that it is a theorem in mathematics and the mathematicians that it is an experimental fact". (iii) Karl Pearson (best known today for the invention of the product-moment correlation coefficient) used his newly developed *Chi Square* test to check how closely a number of empirical distributions of supposedly random errors fitted the bell curve. He found that many of the distributions that had been cited in the literature as fitting the normal curve were actually significantly different from it, and concluded that "the normal curve of error possesses no special fitness for describing errors or deviations such as arise either in observing practice or in nature" (Pearson, 1900).

## 2.2 Randomness of primes

Wells (2005) has discussed the apparent random distribution of prime numbers as follows. The prime numbers are so irregular that is tempting to think of them as some kind of random sequence, in which case it should be possible to use the theory of probability and statistics to study them. The first and most famous application of probability to primes was the Erdös-Kac theorem . The *normal law* states, very roughly, that many distributions in nature behave as if they were the result of tossing a coin many times. Poincare claimed that there is something mysterious about the normal law because mathematicians think it is a law of nature but physicists believe it is a mathematical theorem (Kac, 1959). Recent statistical analyses by Kumar et al (2008), Scafetta *et al.* (2004) and earlier studies by Wolf (1996, 1997) show a new kind of order, namely, selfsimilarity in the spacing intervals of prime numbers.

Despite the huge advances in number theory, many properties of the prime numbers are still unknown, and they appear to us as a random collection of numbers without much structure. In the last few years, some numerical investigations related with the statistical properties of the prime number sequence (Wolf, 1996, 1999; Dahmen *et al*., 2001) have revealed that, apparently, some regularity actually exists in the differences and increments (differences of differences) of consecutive prime numbers (Ares and Castro, 2006). The importance of the study of prime number distribution is summarized by Dahmen *et al* (2001) as follows: The interest in prime numbers is manifold: from a purely mathematical point of view, primes are the building blocks of natural numbers and it comes as no surprise that the study of their properties has attracted and still attracts the attention of some of the most brilliant mathematicians (Zagier, 1977). Primes are also at the base of the RSA—Public Key Encryption System (Rivest *et al*, 1978) and, from a physicist's point of view, besides more direct applications in acoustics (Schroeder, 1990), primes have attracted the attention of quantum "chaologists" due to their intrinsic relations to periodic orbits in dynamical systems (Berry and Keating, 1999).

### 2.3 Fractal fluctuations and statistical analysis

Fractals are the latest development in statistics. The space-time fluctuation pattern in dynamical systems was shown to have a selfsimilar or fractal structure in the 1970s (Mandelbrot, 1975). The larger scale fluctuation consists of smaller scale fluctuations identical in shape to the larger scale. An appreciation of the properties of fractals is changing the most basic ways we analyze and interpret data from experiments and is leading to new insights into understanding physical, chemical, biological, psychological, and social systems. Fractal systems extend over many scales and so cannot be characterized by a single characteristic average number (Liebovitch and Scheurle, 2000). Further, the selfsimilar fluctuations imply long-range space-time correlations or interdependence. Therefore, the Gaussian distribution will not be applicable for description of fractal data sets. However, the bell curve still continues to be used for approximate quantitative characterization of data which are now identified as fractal space-time fluctuations.

### 2.2.1 Power laws and fat tails

Fractals conform to power laws. A power law is a relationship in which one quantity *A* is proportional to another *B* taken to some power *n*; that is, $A \sim B^n$ (Buchanan, 2004). One of the oldest scaling laws in geophysics is the Omori law (Omori, 1895). This law describes the temporal distribution of the number of after-shocks, which occur after a larger earthquake (i.e., main-shock) by a scaling relationship. Richardson (1960) came close to the concept of fractals when he noted that the estimated length of an irregular coastline scales with the length of the measuring unit. Andriani and McKelvey (2007) have given exhaustive references to earliest known work on power law relationships summarized as follows. Pareto (1897) first noticed power laws and fat tails in economics. Cities follow a power law when ranked by population (Auerbach, 1913). Dynamics of earthquakes follow power law (Gutenberg and Richter, 1944) and Zipf (1949) found that a power law applies to word frequencies (Estoup (1916), had earlier found a similar relationship). Mandelbrot (1963) rediscovered them in the 20th century, spurring a small wave of interest in finance (Fama, 1965; Montroll and Shlesinger, 1984). However, the rise of the 'standard' model (Gaussian) of efficient markets, sent power law models into obscurity. This lasted until the 1990s, when the occurrence of catastrophic events, such as the 1987 and 1998 financial crashes, that were difficult to explain with the 'standard' models (Bouchaud *et al*., 1998), re-kindled the fractal model (Mandelbrot and Hudson, 2004).

There are many physical and/or mathematical mechanisms that generate power law distributions and self-similar behavior. Understanding how a mechanism is selected by the microscopic laws constitute an active field of research (Sornette, 2007). Sornette (1995) cites the works of Mandelbrot (1983), Aharony and Feder (1989) and Riste and Sherrington (1991) and states that observation that many natural phenomena have size distributions that are power laws, has been taken as a fundamental indication of an underlying self-similarity. A power law distribution indicates the absence of a characteristic size and as a consequence that there is no upper limit on the size of events. The largest events of a power law distribution completely dominate the underlying physical process; for instance, fluid-driven erosion is dominated by the largest floods and most deformation at plate boundaries takes place through the agency of the largest earthquakes. It is a matter of debate whether power law distributions, which are valid descriptions of the numerous small and intermediate events, can be extrapolated to large events; the largest events are, almost by definition, undersampled.

A power law world is dominated by extreme events ignored in a Gaussian-world. In fact, the fat tails of power law distributions make large extreme events orders-of-magnitude more likely. Theories explaining power laws are also scale-free. This is to say, the same explanation (theory) applies at all levels of analysis (Andriani and McKelvey, 2007).

### 2.2.2 Scale-free theory for power laws with fat, long tails

A scale-free theory for the observed fractal fluctuations in atmospheric flows shows that the observed long-range spatiotemporal correlations are intrinsic to quantumlike chaos governing fluid flows. The model concepts are independent of the exact details such as the chemical, physical, physiological and other properties of the dynamical system and therefore provide a general systems theory applicable to all real world and computed model dynamical systems in nature (Selvam, 1993, 1998, 1999, 2001, 2001a, 2001b, 2002a, 2002b, 2004, 2005, 2007; Selvam *et al.*, 2000). The model is based on the concept that the irregular fractal fluctuations may be visualized to result from the superimposition of an eddy continuum, i.e., a hierarchy of eddy circulations generated at each level by the space-time integration of enclosed small-scale eddy fluctuations. Such a concept of space-time fluctuation averaged distributions *should* follow statistical normal distribution according to *Central Limit Theorem* in traditional Statistical theory (Ruhla, 1992). Also, traditional statistical/mathematical theory predicts that the Gaussian, its Fourier transform and therefore Fourier transform associated power spectrum are the same distributions. The Fourier transform of normal distribution is essentially a normal distribution. A power spectrum is based on the Fourier transform, which expresses the relationship between time (space) domain and frequency domain description of any physical process (Phillips, 2005; Riley, Hobson and Bence, 2006). However, the general systems theory model (Sec. 4) visualises the eddy growth process in successive stages of unit length-step growth with ordered two-way energy feedback between the larger and smaller scale eddies and derives a power law probability distribution $P$ which is close to the Gaussian for small deviations and gives the observed fat, long tail for large fluctuations. Further, the model predicts the power spectrum of the eddy continuum also to follow the power law probability distribution $P$.

In summary, the model predicts the following:

- The eddy continuum consists of an overall logarithmic spiral trajectory with the quasiperiodic *Penrose* tiling pattern for the internal structure.
- The successively larger eddy space-time scales follow the Fibonacci number series.
- The probability distribution $P$ of fractal domains for the $n^{th}$ step of eddy growth is equal to $\tau^{-4n}$ where $\tau$ is the golden mean equal to $(1+\sqrt{5})/2$ ($\approx 1.618$). The eddy growth step $n$ represents the *normalized deviation t* in traditional statistical theory. The *normalized deviation t* represents the departure of the variable from the mean in terms of the standard deviation of the distribution assumed to follow *normal distribution* characteristics for many real world space-time events. There is progressive decrease in the probability of occurrence of events with increase in corresponding *normalized deviation t*. Space-time events with *normalized deviation t* greater than 2 occur with a probability of 5 percent or less and may be categorized as extreme events associated in general with widespread (space-time) damage and loss. The model predicted probability distribution *P* is close to the statistical normal distribution for *t* values less than 2 and greater than normal distribution for *t* more than 2, thereby giving a *fat, long tail*. There is non-zero probability of occurrence of very large events.

- The inverse of probability distribution *P*, namely, $\tau^{4n}$ represents the relative eddy energy flux in the large eddy fractal (small scale fine structure) domain. There is progressive decrease in the probability of occurrence of successive stages of eddy growth associated with progressively larger domains of fractal (small scale fine structure) eddy energy flux and at sufficiently large growth stage trigger catastrophic extreme events such as heavy rainfall, stock market crashes, traffic jams, etc., in real world situations.

- The power spectrum of fractal fluctuations also follows the same distribution *P* as for the distribution of fractal fluctuations. The square of the eddy amplitude (variance) represents the eddy energy and therefore the eddy probability density *P*. Such a result that the additive amplitudes of eddies when squared represent probabilities, is exhibited by the sub-atomic dynamics of quantum systems such as the electron or proton (Maddox, 1988, 1993; Rae, 1988). The phase spectrum is the same as the variance spectrum, a characteristic of quantum systems identified as '*Berry's phase*'. Fractal fluctuations are signatures of quantumlike chaos in dynamical systems.

- The *fine structure constant* for spectrum of fractal fluctuations is a function of the *golden mean* and is analogous to that of atomic spectra equal to about 1/137.

- The universal algorithm for self-organized criticality is expressed in terms of the universal *Feigenbaum*'s constants (Feigenbaum, 1980) *a* and *d* as $2a^2 = \pi d$ where the fractional volume intermittency of occurrence $\pi d$ contributes to the total variance $2a^2$ of fractal structures. The *Feigenbaum*'s constants are expressed as functions of the *golden mean*. The probability distribution *P* of fractal domains is also expressed in terms of the Feigenbaum's constants *a* and *d*. The details of the model are summarized in the Sec. 4.

## 3. Deterministic chaos and fractal fluctuations in computed model dynamical systems

The continuum real number field (infinite number of decimals between any two integers) represented as Cartesian co-ordinates (Mathews, 1961; Stewart and Tall, 1990; Devlin, 1997; Stewart, 1996, 1998) is the basic computational tool in the simulation and prediction of the continuum dynamics of real world dynamical systems such as fluid flows, stock market price fluctuations, heart beat patterns, etc. Till the late 1970s, mathematical models were based on Newtonian continuum dynamics with implicit assumption of linearity in the rate of change with respect to (w. r. t) time or space of the dynamical variable under consideration. The traditional mathematical model equations were of the form

$$X_{n+1} = X_n + \left(\frac{\mathrm{d}X}{\mathrm{d}t}\right)_n \mathrm{d}t \tag{1}$$

Constant value was assumed for the rate of change $(\mathrm{d}X/\mathrm{d}t)_n$ of the variable $X_n$ at computational step *n* and infinitesimally small time or space intervals *dt*. Eq. (1) will be linear and can be solved analytically provided the rate of change $(dX/dt)_n$ is constant. However, dynamical systems in nature exhibit irregular (*fractal*) fluctuations on all space and time scales and therefore the assumption of constant rate of change fails and Eq. (1) does not have analytical solution. Numerical solutions are then obtained for discrete (finite) space-time intervals such that the continuum dynamics of Eq. (1) is now computed as discrete dynamics given by

$$X_{n+1} = X_n + \left(\frac{\Delta X}{\Delta t}\right)_n \Delta t \qquad (2)$$

Numerical solutions obtained using Eq. (2), which is basically a numerical integration procedure, involve iterative computations with feedback and amplification of round-off error of real number finite precision arithmetic. The Eq. (2) also represents the relationship between continuum number field and embedded discrete (finite) number fields. Numerical solutions for non-linear dynamical systems represented by Eq. (2) are sensitively dependent on initial conditions and give apparently chaotic solutions, identified as *deterministic chaos*. *Deterministic chaos* therefore characterises the evolution of discrete (finite) structures from the underlying continuum number field.

Historically, sensitive dependence on initial conditions of non-linear dynamical systems was identified nearly a century ago by *Poincare* (Poincare, 1892) in his study of three-body problem, namely the sun, earth and the moon. Non-linear dynamics remained a neglected area of research till the advent of electronic computers in the late 1950s. Lorenz, in 1963 showed that numerical solutions of a simple model of atmospheric flows exhibited sensitive dependence on initial conditions implying loss of predictability of the future state of the system. The traditional non-linear dynamical system defined by Eq. (2) is commonly used in all branches of science and other areas of human interest. *Non-linear dynamics and chaos* soon (by 1980s) became a multidisciplinary field of intensive research (Gleick, 1987). Sensitive dependence on initial conditions implies long-range space-time correlations. The observed irregular fluctuations of real world dynamical systems also exhibit such non-local connections manifested as *fractal* or *self-similar* geometry to the space-time evolution. The universal symmetry of *self-similarity* ubiquitous to dynamical systems in nature is now identified as *self-organized criticality* (Bak, Tang and Wiesenfeld, 1988). A symmetry of some figure or pattern is a transformation that leaves the figure invariant, in the sense that, taken as a whole it looks the same after the transformation as it did before, although individual points of the figure may be moved by the transformation (Devlin, 1997). *Self-similar* structures have internal geometrical structure, which resemble the whole. The space-time organization of a hierarchy of *self-similar* space-time structures is common to real world as well as the numerical models (Eq. 2) used for simulation. A substratum of continuum fluctuations self-organizes to generate the observed unique hierarchical structures both in real world and the continuum number field used as the tool for simulation. A *cell dynamical system model* developed by the author (Selvam, 1990; Selvam and Fadnavis, 1998; 1999a, b) for turbulent fluid flows shows that *self-similar* (*fractal*) space-time fluctuations exhibited by real world and numerical models of dynamical systems are signatures of quantum-like chaos. The model concepts are independent of the exact details, such as, the chemical, physical, physiological, etc., properties of the dynamical systems and therefore provide a general systems theory (Peacocke, 1989; Klir, 1993; Jean, 1994) applicable for all dynamical systems in nature. The model concepts are applicable to the emergence of unique prime number spectrum from the underlying substratum of continuum real number field.

Wolf (1999) cites the reported numerous links between number theory and physics as follows: see e.g. two books (Luck *et al.*, 1990; Waldschmidt *et al.*, 1992). Very well known are applications of number theory in chaos, both classical and quantum. As an example the Fibonnaci numbers can be mentioned: there is an ubiquity of places in the theory of chaos, where they appear (see Schuster, 1989). Other papers where some mathematical facts about primes were applied to the study of quantum chaos can be found in Gutzwiller (1992, 2008), Berry (1993), Aurich (1993, 1994), Sarnak (1995). Wolf (1989) investigated the multifractality of primes and Wolf (1997) found $1/f$ noise in the distribution of primes. Recent studies indicate a close association between *number theory* in mathematics, in particular, the distribution of *prime numbers* and the chaotic orbits of excited quantum systems such as the hydrogen atom (Keating, 1990; Cipra, 1996; Berry and Keating, 1999; Klarreich, 2000). Mathematical studies also indicate that *Cantorian fractal* space-time characterises quantum systems (Ord, 1983; Nottale, 1989; El Naschie, 1993, 2008). The *fractal* fluctuations exhibited by prime number distribution and microscopic quantum systems belong to the newly identified science of *non-linear dynamics and chaos*. Quantification of the apparently irregular (chaotic) *fractal* fluctuations will help compute (predict) the space-time evolution of the fluctuations. The *general systems theory model* concepts described below (Sec. 4) provide a theory for unique quantification of the observed *fractal* fluctuations in terms of the universal inverse power-law form incorporating the *golden mean*.

## 4. A general systems theory for fractal fluctuations

The fractal space-time fluctuations of dynamical systems may be visualized to result from the superimposition of an ensemble of eddies (sine waves), namely an eddy continuum. The relationship between large and small eddy circulation parameters are obtained on the basis of Townsend's (1956) concept that large eddies are envelopes enclosing turbulent eddy (small-scale) fluctuations (Fig. 2).

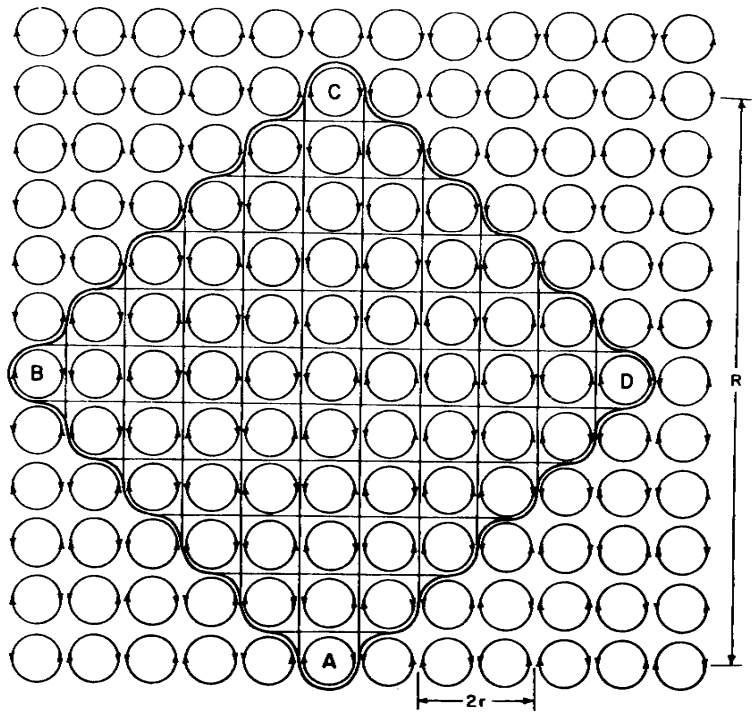

Fig. 2: Visualisation of the formation of large eddy (ABCD) as envelope enclosing smaller scale eddies. By analogy, the continuum number field domain (Cartesian co-ordinates) may also be obtained from successive integration of enclosed finite number field domains.

The relationship between root mean square (r. m. s.) circulation speeds $W$ and $w_*$ respectively of large and turbulent eddies of respective radii $R$ and $r$ is then given as

$$W^2 = \frac{2}{\pi} \frac{r}{R} w_*^2 \tag{3}$$

The dynamical evolution of space-time fractal structures is quantified in terms of ordered energy flow between fluctuations of all scales in Eq. (3), because the square of the eddy circulation speed represents the eddy energy (kinetic). A hierarchical continuum of eddies is generated by the integration of successively larger enclosed turbulent eddy circulations. Such a concept of space-time fluctuation averaged distributions *should* follow statistical normal distribution according to *Central Limit Theorem* in traditional Statistical theory (Ruhla, 1992). Also, traditional statistical/mathematical theory predicts that the Gaussian, its Fourier transform and therefore Fourier transform associated power spectrum are the same distributions. However, the general systems theory (Selvam, 1998, 1999, 2001a, 2001b, 2002a, 2002b, 2004, 2005, 2007; Selvam *et al.*, 2000) visualises the eddy growth process in successive stages of unit length-step growth with ordered two-way energy feedback between the larger and smaller scale eddies and derives a power law probability distribution $P$ which is close to the Gaussian for small deviations and gives the observed fat, long tail for large fluctuations. Further, the model predicts the power spectrum of the eddy continuum also to follow the power law probability distribution $P$. Therefore the additive amplitudes of the eddies when squared (variance), represent the probability distribution similar to the subatomic dynamics of quantum systems such as the electron or photon. Fractal fluctuations therefore exhibit quantumlike chaos.

The above-described analogy of quantumlike mechanics for dynamical systems is similar to the concept of a subquantum level of fluctuations whose space-time organization gives rise to the observed manifestation of subatomic phenomena, i.e., quantum systems as order out of chaos phenomena (Grossing, 1989).

### 4.1 Quasicrystalline structure of the eddy continuum

The turbulent eddy circulation speed and radius increase with the progressive growth of the large eddy (Selvam, 1990, 2007). The successively larger turbulent fluctuations, which form the internal structure of the growing large eddy, may be computed (Eq. 3) as

$$w_*^2 = \frac{\pi}{2} \frac{R}{dR} W^2 \tag{4}$$

During each length step growth $dR$, the small-scale energizing perturbation $W_n$ at the $n^{\text{th}}$ instant generates the large-scale perturbation $W_{n+1}$ of radius $R$ where $R = \sum_{1}^{n} dR$ since successive length-scale doubling gives rise to $R$. Eq. (4) may be written in terms of the successive turbulent circulation speeds $W_n$ and $W_{n+1}$ as

$$W_{n+1}^2 = \frac{\pi}{2} \frac{R}{dR} W_n^2 \tag{5}$$

The angular turning $d\theta$ inherent to eddy circulation for each length step growth is equal to $dR/R$. The perturbation $dR$ is generated by the small-scale acceleration $W_n$ at any instant $n$ and therefore $dR = W_n$. Starting with the unit value for $dR$ the successive $W_n$, $W_{n+1}$, $R$, and $d\theta$ values are computed from Eq. 5 and are given in Table 1.

Table 1. The computed spatial growth of the strange-attractor design traced by the macro-scale dynamical system of atmospheric flows as shown in Fig. 3.

| $R$ | $W_n$ | d$R$ | d$\theta$ | $Wn+1$ | $\theta$ |
|-------|--------|--------|--------|--------|--------|
| 1.000 | 1.000 | 1.000 | 1.000 | 1.254 | 1.000 |
| 2.000 | 1.254 | 1.254 | 0.627 | 1.985 | 1.627 |
| 3.254 | 1.985 | 1.985 | 0.610 | 3.186 | 2.237 |
| 5.239 | 3.186 | 3.186 | 0.608 | 5.121 | 2.845 |
| 8.425 | 5.121 | 5.121 | 0.608 | 8.234 | 3.453 |
| 13.546 | 8.234 | 8.234 | 0.608 | 13.239 | 4.061 |
| 21.780 | 13.239 | 13.239 | 0.608 | 21.286 | 4.669 |
| 35.019 | 21.286 | 21.286 | 0.608 | 34.225 | 5.277 |
| 56.305 | 34.225 | 34.225 | 0.608 | 55.029 | 5.885 |
| 90.530 | 55.029 | 55.029 | 0.608 | 88.479 | 6.493 |

It is seen that the successive values of the circulation speed $W$ and radius $R$ of the growing turbulent eddy follow the Fibonacci mathematical number series such that $R_{n+1}=R_n+R_{n-1}$ and $R_{n+1}/R_n$ is equal to the golden mean $\tau$, which is equal to $[(1 + \sqrt{5})/2] \cong 1.618$. Further, the successive $W$ and $R$ values form the geometrical progression $R_0(1+\tau+\tau^2+\tau^3+\tau^4+ ....)$ where $R_0$ is the initial value of the turbulent eddy radius.

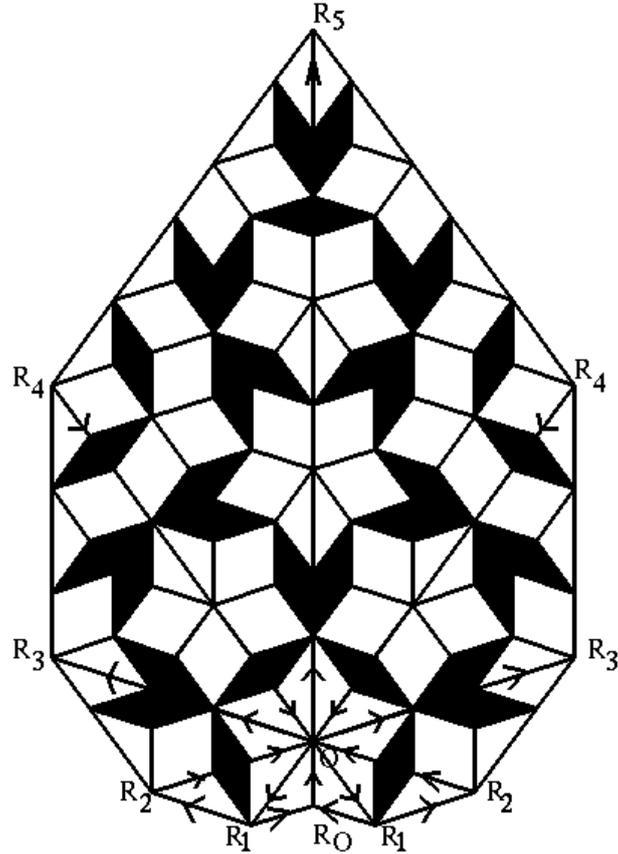

Fig. 3: The quasiperiodic *Penrose* tiling pattern with *five-fold* symmetry traced by the small eddy circulations internal to dominant large eddy circulation in turbulent fluid flows

Turbulent eddy growth from primary perturbation $OR_O$ starting from the origin O (Fig. 3) gives rise to compensating return circulations $OR_1R_2$ on either side of $OR_O$, thereby generating the large eddy radius $OR_1$ such that $OR_1/OR_O=\tau$ and $R_OOR_1=\pi/5=R_OR_1O$.

Therefore, short-range circulation balance requirements generate successively larger circulation patterns with precise geometry that is governed by the *Fibonacci* mathematical number series, which is identified as a signature of the universal period doubling route to chaos in fluid flows, in particular atmospheric flows. It is seen from Fig. 3 that five such successive length step growths give successively increasing radii $OR_1$, $OR_2$, $OR_3$, $OR_4$ and $OR_5$ tracing out one complete vortex-roll circulation such that the scale ratio $OR_5/OR_O$ is equal to $\tau^5 = 11.1$. The envelope $R_1R_2R_3R_4R_5$ (Fig. 3) of a dominant large eddy (or vortex roll) is found to fit the logarithmic spiral $R = R_0 e^{b\theta}$ where $R_0 = OR_O$, $b = \tan\delta$ with $\delta$ the crossing angle equal to $\pi/5$, and the angular turning $\theta$ for each length step growth is equal to $\pi/5$. The successively larger eddy radii may be subdivided again in the *golden mean* ratio. The internal structure of large-eddy circulations is therefore made up of balanced small-scale circulations tracing out the well-known quasi-periodic *Penrose* tiling pattern identified as the quasi-crystalline structure in condensed matter physics. A complete description of the atmospheric flow field is given by the quasi-periodic cycles with *Fibonacci* winding numbers.

## 4.2 Model predictions

The model predictions (Selvam, 1990, 2005, 2007; Selvam and Fadnavis, 1998) are

### 4.2.1 Quasiperiodic Penrose tiling pattern

Atmospheric flows trace an overall logarithmic spiral trajectory $OR_OR_1R_2R_3R_4R_5$ simultaneously in clockwise and anti-clockwise directions with the quasi-periodic *Penrose tiling pattern* (Steinhardt, 1997) for the internal structure shown in Fig. 3.

   The spiral flow structure can be visualized as an eddy continuum generated by successive length step growths $OR_O$, $OR_1$, $OR_2$, $OR_3$,....respectively equal to $R_1$, $R_2$, $R_3$,....which follow *Fibonacci* mathematical series such that $R_{n+1} = R_n + R_{n-1}$ and $R_{n+1}/R_n = \tau$ where $\tau$ is the *golden mean* equal to $(1+\sqrt{5})/2$ ($\approx 1.618$). Considering a normalized length step equal to 1 for the last stage of eddy growth, the successively decreasing radial length steps can be expressed as 1, $1/\tau$, $1/\tau^2$, $1/\tau^3$, ......The normalized eddy continuum comprises of fluctuation length scales 1, $1/\tau$, $1/\tau^2$, ........ The probability of occurrence is equal to $1/\tau$ and $1/\tau^2$ respectively for eddy length scale $1/\tau$ in any one or both rotational (clockwise and anti-clockwise) directions. Eddy fluctuation length of amplitude $1/\tau$ has a probability of occurrence equal to $1/\tau^2$ in both rotational directions, i.e., the square of eddy amplitude represents the probability of occurrence in the eddy continuum. Similar result is observed in the subatomic dynamics of quantum systems which are visualized to consist of the superimposition of eddy fluctuations in wave trains (eddy continuum).

### 4.2.2 Eddy continuum

Conventional continuous periodogram power spectral analyses of such spiral trajectories in Fig. 3 $(R_oR_1R_2R_3R_4R_5)$ will reveal a continuum of periodicities with progressive increase $d\theta$ in phase angle $\theta$ (theta) as shown in Fig. 4.

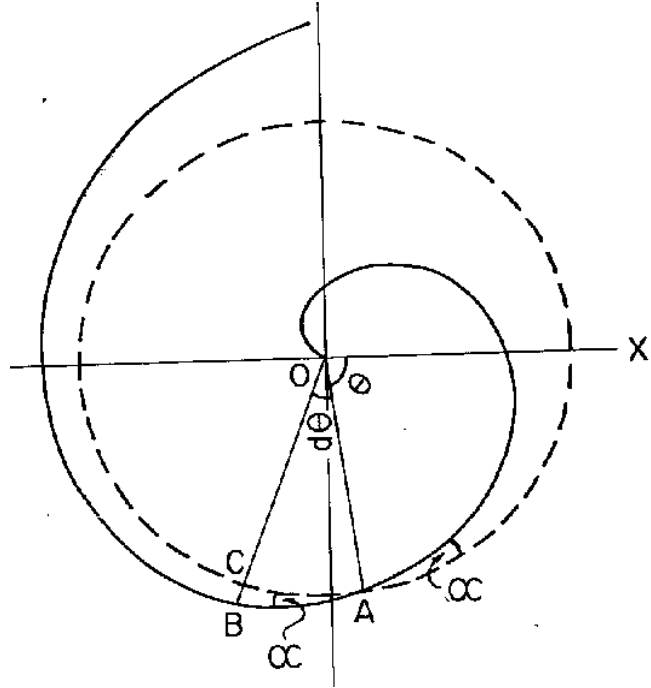

Fig. 4: The equiangular logarithmic spiral given by $(R/r) =$ $e^{\alpha\theta}$ where $\alpha$ and $\theta$ are each equal to $1/z$ for each length step growth. The eddy length scale ratio $z$ is equal to $R/r$. The crossing angle $\alpha$ is equal to the small increment $d\theta$ in the phase angle $\theta$ Traditional power spectrum analysis will resolve such a spiral flow trajectory as a continuum of eddies with progressive increase $d\theta$ in phase angle $\theta$.

### 4.2.3 Dominant eddies

The broadband power spectrum will have embedded dominant wavebands ($R_0OR_1$, $R_1OR_2$, $R_2OR_3$, $R_3OR_4$, $R_4OR_5$, etc.) the bandwidth increasing with period length (Fig. 3). The peak periods $E_n$ in the dominant wavebands is be given by the relation

$$E_n = T_s(2 + \tau)\tau^n \qquad (6)$$

where $\tau$ is the *golden mean* equal to $(1+\sqrt{5})/2$ (approximately equal to 1.618) and $T_s$ , the primary perturbation time period, for example, is the annual cycle (summer to winter) of solar heating in a study of atmospheric interannual variability. The peak periods $E_n$ are superimposed on a continuum background. For example, the most striking feature in climate variability on all time scales is the presence of sharp peaks superimposed on a continuous background (Ghil, 1994). In the case of prime number frequency distribution at unit number intervals, the model predicted (Eq. 6) dominant peak periodicities are *2.2, 3.6, 5.8, 9.5, 15.3, 24.8, 40.1,* and *64.9* unit number spacing intervals for values of $n$ ranging from -1 to 6.

### 4.2.4 Berry's phase in quantum systems

The ratio $r/R$ also represents the increment $d\theta$ in phase angle $\theta$ (Eq. 5 and Fig. 4) and therefore the phase angle $\theta$ represents the variance (Selvam, 1990, 2007). Hence, when the logarithmic spiral is resolved as an eddy continuum in conventional spectral analysis, the increment in wavelength is concomitant with increase in phase. The angular turning, in turn, is directly proportional to the variance (Eq. 5). Such a result that increments in wavelength and phase angle are related is observed in quantum systems and has been named *'Berry's* *phase*' (Berry, 1988). The relationship of angular turning of the spiral to intensity of fluctuations is seen in the tight coiling of the hurricane spiral cloud systems.

*4.2.5 Logarithmic spiral pattern underlying fractal fluctuations*

The overall logarithmic spiral flow structure is given by the relation

$$W = \frac{w_*}{k} \ln z \tag{7}$$

where the constant $k$ is the steady state fractional volume dilution of large eddy by inherent turbulent eddy fluctuations. The constant $k$ is equal to $k = \frac{w_* r}{WR} = \frac{1}{\tau^2} \approx 0.382$ is identified as the universal constant for deterministic chaos in fluid flows (Selvam, 1990, 2007). Since $k$ is less than half, the mixing with environmental air does not erase the signature of the dominant large eddy, but helps to retain its identity as a stable self-sustaining *soliton-like* structure. The mixing of environmental air assists in the upward and outward growth of the large eddy. The steady state emergence of fractal structures is therefore equal to

$$\frac{1}{k} \approx 2.62 \tag{8}$$

   Logarithmic wind profile relationship such as Eq. 7 is a long-established (observational) feature of atmospheric flows in the boundary layer, the constant $k$, called the *Von Karman*'s constant has the value equal to 0.38 as determined from observations (Wallace and Hobbs, 1977). In Eq. 7, $W$ represents the standard deviation of eddy fluctuations, since $W$ is computed as the instantaneous r.m.s. (root mean square) eddy perturbation amplitude with reference to the earlier step of eddy growth. For two successive stages of eddy growth starting from primary perturbation $w_*$ the ratio of the standard deviations $W_{n+1}$ and $W_n$ is given from Eq. 7 as *(n+1)/n*. Denoting by $\sigma$ the standard deviation of eddy fluctuations at the reference level ($n=1$) the standard deviations of eddy fluctuations for successive stages of eddy growth are given as integer multiple of $\sigma$, i.e., $\sigma$, $2\sigma$, $3\sigma$, etc. and correspond respectively to

$$\text{statistical normalised standard deviation} \quad t = 0, 1, 2, 3, .... \tag{9}$$

# 5. Universal Feigenbaum's constants and probability density distribution function for fractal fluctuations

Selvam (1993, 2007) has shown that Eq. (3) represents the universal algorithm for deterministic chaos in dynamical systems and is expressed in terms of the universal *Feigenbaum*'s (1980) *constants a* and *d* as follows. The successive length step growths generating the eddy continuum $OR_O R_1 R_2 R_3 R_4 R_5$ (Fig. 3) analogous to the period doubling route to chaos (growth) is initiated and sustained by the turbulent (fine scale) eddy acceleration $w_*$, which then propagates by the inherent property of inertia of the medium of propagation. Therefore, the statistical parameters *mean*, *variance*, *skewness* and *kurtosis* of the perturbation field in the medium of propagation are given by $w_*, w_*^2, w_*^3$ and $w_*^4$ respectively. The associated dynamics of the perturbation field can be described by the following parameters. The perturbation speed $w_*$ (motion) per second (unit time) sustained by its inertia represents the mass, $w_*^2$ the acceleration or force, $w_*^3$ the angular momentum or potential energy, and $w_*^4$ the spin angular momentum, since an eddy motion has an inherent curvature to its trajectory.

It is shown that *Feigenbaum's* constant $a$ is equal to (Selvam, 1993, 2007)

$$a = \frac{W_2 R_2}{W_1 R_1} \qquad (10)$$

In Eq. (10) the subscripts 1 and 2 refer to two successive stages of eddy growth. *Feigenbaum's* constant $a$ as defined above represents the steady state emergence of fractional *Euclidean* structures. Considering dynamical eddy growth processes, *Feigenbaum's* constant $a$ also represents the steady state fractional outward mass dispersion rate and $a^2$ represents the energy flux into the environment generated by the persistent primary perturbation $W_1$. Considering both clockwise and counterclockwise rotations, the total energy flux into the environment is equal to $2a^2$. In statistical terminology, $2a^2$ represents the variance of fractal structures for both clockwise and counterclockwise rotation directions.

The steady state emergence of fractal structures in fluid flows is equal to $1/k$ ($=\tau^2$) (Eq. 8) and therefore the *Feigenbaum's constant $a$* is equal to

$$a = \tau^2 = \frac{1}{k} \approx 2.62 \qquad (11)$$

The probability of occurrence $P_{tot}$ of fractal domain $W_1 R_1$ in the total larger eddy domain $W_n R_n$ in any (irrespective of positive or negative) direction is equal to

$$P_{tot} = \frac{W_1 R_1}{W_n R_n} = \tau^{-2n}$$

Therefore the probability $P$ of occurrence of fractal domain $W_1 R_1$ in the total larger eddy domain $W_n R_n$ in any one direction (either positive or negative) is equal to

$$P = \left( \frac{W_1 R_1}{W_n R_n} \right)^{2n} = \tau^{-4n} \qquad (12)$$

The *Feigenbaum's* constant $d$ is shown to be equal to (Selvam, 1993, 2007)

$$d = \frac{W_2^4 R_2^3}{W_1^4 R_1^3} \qquad (13)$$

Eq. (13) represents the fractional volume intermittency of occurrence of fractal structures for each length step growth. *Feigenbaum's* constant $d$ also represents the relative spin angular momentum of the growing large eddy structures as explained earlier.

Eq. (3) may now be written as

$$2 \frac{W^2 R^2}{w_*^2 (dR)^2} = \pi \frac{W^4 R^3}{w_*^4 (dR)^3} \qquad (14)$$

In Eq. (14) $dR$ equal to $r$ represents the incremental growth in radius for each length step growth, i.e., $r$ relates to the earlier stage of eddy growth.

The Feigenbaum's constant $d$ represented by $R/r$ is equal to

$$d = \frac{W^4 R^3}{w_*^4 r^3} \qquad (15)$$

For two successive stages of eddy growth

$$d = \frac{W_2^4 R_2^3}{W_1^4 R_1^3} \tag{16}$$

From Eq. (3)

$$W_1^2 = \frac{2}{\pi} \frac{r}{R_1} w_*^2$$

$$W_2^2 = \frac{2}{\pi} \frac{r}{R_2} w_*^2 \tag{17}$$

Therefore

$$\frac{W_2^2}{W_1^2} = \frac{R_1}{R_2} \tag{18}$$

Substituting in Eq. (16)

$$d = \frac{W_2^4 R_2^3}{W_1^4 R_1^3} = \frac{W_2^2}{W_1^2} \frac{W_2^2 R_2^3}{W_1^2 R_1^3} = \frac{R_1}{R_2} \frac{W_2^2 R_2^3}{W_1^2 R_1^3} = \frac{W_2^2 R_2^2}{W_1^2 R_1^2} \tag{19}$$

The Feigenbaum's constant $d$ represents the scale ratio $R_2/R_1$ and the inverse of the Feigenbaum's constant $d$ equal to $R_1/R_2$ represents the probability $(Prob)_1$ of occurrence of length scale $R_1$ in the total fluctuation length domain $R_2$ for the first eddy growth step as given in the following

$$(Prob)_1 = \frac{R_1}{R_2} = \frac{1}{d} = \frac{W_1^2 R_1^2}{W_2^2 R_2^2} = \tau^{-4} \tag{20}$$

In general for the $n^{th}$ eddy growth step, the probability $(Prob)_n$ of occurrence of length scale $R_1$ in the total fluctuation length domain $R_n$ is given as

$$(Prob)_n = \frac{R_1}{R_n} = \frac{W_1^2 R_1^2}{W_n^2 R_n^2} = \tau^{-4n} \tag{21}$$

The above equation for probability $(Prob)_n$ also represents, for the $n^{th}$ eddy growth step, the following statistical and dynamical quantities of the growing large eddy with respect to the initial perturbation domain: (i) the statistical relative variance of fractal structures, (ii) probability of occurrence of fractal domain in either positive or negative direction, and (iii) the inverse of $(Prob)_n$ represents the normalized fractal (fine scale) energy flux in the overall large scale eddy domain. Large scale energy flux therefore occurs not in bulk, but in organized internal fine scale circulation structures identified as fractals.

Substituting the *Feigenbaum's constants a* and *d* defined above (Eqs. 10 and 13), Eq. (14) can be written as

$$2a^2 = \pi d \tag{22}$$

In Eq. (22) $\pi d$, the relative volume intermittency of occurrence contributes to the total variance $2a^2$ of fractal structures.

In terms of eddy dynamics, the above equation states that during each length step growth, the energy flux into the environment equal to $2a^2$ contributes to generate relative spin angular momentum equal to $\pi d$ of the growing fractal structures. Each length step growth is therefore associated with a factor of $2a^2$ equal to $2\tau^4$ ( $\cong 13.7082$) increase in energy flux in the associated fractal domain. Ten such length step growths results in the formation of robust (self-sustaining) dominant bidirectional large eddy circulation $OR_OR_1R_2R_3R_4R_5$ (Fig. 3) associated with a factor of $20a^2$ equal to $137.08203$ increase in eddy energy flux. This non-dimensional constant factor characterizing successive dominant eddy energy increments is analogous to the *fine structure* constant $\propto^{-1}$ (Ford, 1968) observed in atomic spectra, where the spacing (energy) intervals between adjacent spectral lines is proportional to the non-dimensional *fine structure* constant equal to approximately $1/137$. Further, the probability of $n$th length step eddy growth is given by $a^{-2n}$ ($\cong 6.854^{-n}$) while the associated increase in eddy energy flux into the environment is equal to $a^{2n}$ ($\cong 6.854^n$). Extreme events occur for large number of length step growths $n$ with small probability of occurrence and are associated with large energy release in the fractal domain. Each length step growth is associated with one-tenth of *fine structure constant* energy increment equal to $2a^2$ ($\propto^{-1}/10 \cong 13.7082$) for bidirectional eddy circulation, or equal to one-twentieth of *fine structure constant* energy increment equal to $a^2$ ($\propto^{-1}/20 \cong 6.854$) in any one direction, i.e., positive or negative. The energy increase between two successive eddy length step growths may be expressed as a function of $(a^2)^2$, i.e., proportional to the square of the *fine structure constant* $\propto^{-1}$. In the spectra of many atoms, what appears with coarse observations to be a single spectral line proves, with finer observation, to be a group of two or more closely spaced lines. The spacing of these fine-structure lines relative to the coarse spacing in the spectrum is proportional to the square of *fine structure constant*, for which reason this combination is called the *fine-structure constant*. We now know that the significance of the *fine-structure constant* goes beyond atomic spectra (Ford, 1968). The atomic spectra therefore exhibit fractal-like structure within structure for the spectral lines.

## 5.1 Same inverse power law for probability distribution and power spectra of fractal fluctuations

The power, i.e. variance spectra of fluctuations in fluid flows can now be quantified in terms of universal *Feigenbaum's constant a* as follows.

The steady state emergence of fractal structures is equal to the *Feigenbaum's constant a* (Eqs. 10 and 11). The relative variance of fractal structure which also represents the probability $P$ of occurrence of bidirectional fractal domain for each length step growth is then equal to $1/a^2$. The normalized variance $\frac{1}{a^{2n}}$ will now represent the statistical probability density for the $n$th step growth according to model predicted quantumlike mechanics for fluid flows. Model predicted probability density values $P$ are computed as

$$P = \frac{1}{a^{2n}} = \tau^{-4n} \qquad (23)$$

or

$$P = \tau^{-4t} \qquad (24)$$

In Eq. (24) $t$ is the normalized standard deviation (Eq. 9).
The normalized variance and therefore the statistical probability distribution is represented by (from Eq. 12)

$$P = a^{-2t} \qquad (25)$$

In Eq. (25) $P$ is the probability density corresponding to normalized standard deviation $t$. The probability density distribution of fractal fluctuations (Eq. 21) is therefore the same as variance spectrum (Eq. 25) of fractal fluctuations.

The graph of $P$ versus $t$ will represent the power spectrum. The slope $S$ of the power spectrum is equal to

$$S = \frac{\mathrm{d}P}{\mathrm{d}t} \approx -P \qquad (26)$$

The power spectrum therefore follows inverse power law form, the slope decreasing with increase in $t$. Increase in $t$ corresponds to large eddies (low frequencies) and is consistent with observed decrease in slope at low frequencies in dynamical systems.

## 5.2 Inverse power law for fractal fluctuations close to Gaussian distribution

The steady state emergence of fractal structures for each length step growth for any one direction of rotation (either clockwise or anticlockwise) is equal to

$$\frac{a}{2} = \frac{\tau^2}{2}$$

since the corresponding value for both direction is equal to $a$ (Eqs. 10 and 11).

The emerging fractal space-time structures have moment coefficient of kurtosis given by the fourth moment equal to

$$\left(\frac{\tau^2}{2}\right)^4 = \frac{\tau^8}{16} = 2.9356 \approx 3$$

The moment coefficient of skewness for the fractal space-time structures is equal to zero for the symmetric eddy circulations. Moment coefficient of kurtosis equal to 3 and moment coefficient of skewness equal to *zero* characterize the statistical normal distribution. The model predicted power law distribution for fractal fluctuations is close to the Gaussian distribution.

## 5.3 Fat long tail for probability distribution of fractal fluctuations

The model predicted $P$ values corresponding to normalized deviation $t$ values less than 2 are slightly less than the corresponding statistical normal distribution values while the $P$ values are noticeably larger for normalized deviation $t$ values greater than 2 (Table 2 and Fig. 5) and may explain the reported *fat tail* for probability distributions of various physical parameters (Buchanan, 2004). The model predicted $P$ values plotted on a linear scale (Y-axis) shows close agreement with the corresponding statistical normal probability values as seen in Fig.5 (left side). The model predicted $P$ values plotted on a logarithmic scale (Y-axis) shows *fat tail* distribution for normalized deviation $t$ values greater than 2 as seen in Fig.5 (right side).

Table 2: Model predicted and statistical normal probability density distributions

| growth step | normalized deviation | cumulative probability densities (%) | |
|---|---|---|---|
| $n$ | $t$ | model predicted $P = \tau^{-4t}$ | statistical normal distribution |
| 1 | 1 | 14.5898 | 15.8655 |
| 2 | 2 | 2.1286 | 2.2750 |
| 3 | 3 | 0.3106 | 0.1350 |
| 4 | 4 | 0.0453 | 0.0032 |
| 5 | 5 | 0.0066 | $\approx 0.0$ |

**comparison of statistical normal distribution and
computed (theoretical) probability density distribution**

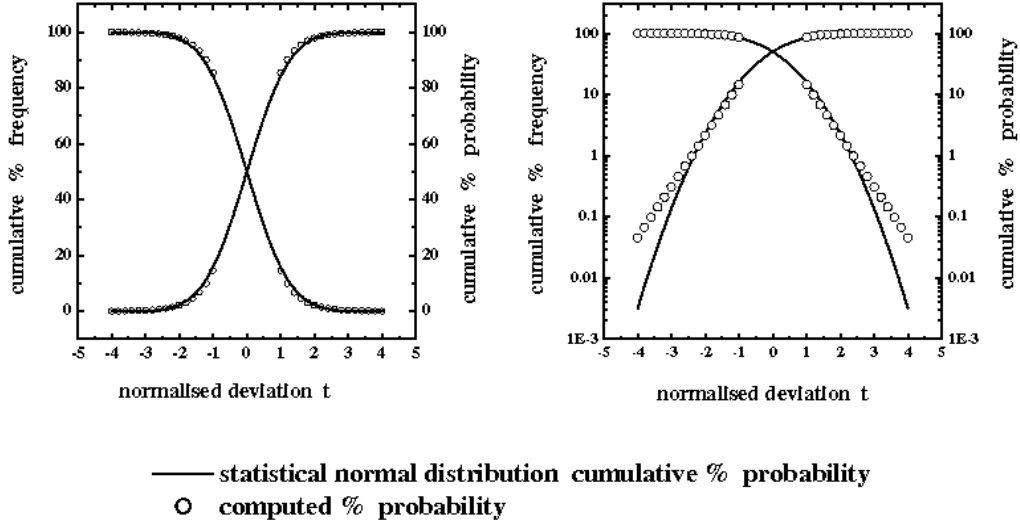

———— **statistical normal distribution cumulative % probability**
O    **computed % probability**

Fig. 5: Comparison of statistical normal distribution and computed (theoretical) probability density distribution. The same figure is plotted on the right side with logarithmic scale for the probability axis (Y-axis) to show clearly that for normalized deviation $t$ values greater than 2 the computed probability densities are greater than the corresponding statistical normal distribution values.

## 5.4 Power spectra of fractal fluctuations

It was shown at Sec. 5.1 above that the same inverse power law represents probability density distribution (Eq. 21) and power spectra (Eq. 25) of fractal fluctuations. Therefore, the conventional power spectrum plotted as the variance versus the frequency in log-log scale will now represent the eddy probability density on logarithmic scale versus the standard deviation of the eddy fluctuations on linear scale since the logarithm of the eddy wavelength represents the standard deviation, i.e., the r. m. s. value of eddy fluctuations (Eq. 7). The r. m. s. value of eddy fluctuations can be represented in terms of probability density distribution as follows. A normalized standard deviation $t=0$ corresponds to cumulative percentage probability density equal to 50 for the mean value of the distribution. Since the logarithm of the wavelength represents the r. m. s. value of eddy fluctuations the normalized standard deviation $t$ is defined for the eddy energy as

$$t = \frac{\log L}{\log T_{50}} - 1 \tag{27}$$

In Eq. (27) $L$ is the time period (or wavelength) and $T_{50}$ is the period up to which the cumulative percentage contribution to total variance is equal to 50 and $t = 0$. Log $T_{50}$ also represents the mean value for the r. m. s. eddy fluctuations and is consistent with the concept of the mean level represented by r. m. s. eddy fluctuations. Spectra of time series of meteorological parameters when plotted as cumulative percentage contribution to total variance versus $t$ have been shown to follow closely the model predicted universal spectrum (Selvam and Fadnavis, 1998) which is identified as a signature of quantumlike chaos.

The period $T_{50}$ up to which the cumulative percentage contribution to total variance is equal to 50 is computed from model concepts as follows. The power spectrum, when plotted as normalized standard deviation $t$ (Eq. 9) versus cumulative percentage contribution to total variance represents the probability density distribution (Sec. 5.1), i.e. the variance represents the probability density. The normalized standard deviation $t$ value 0 corresponds to cumulative percentage probability density $P$ equal to 50, same as for statistical normal distribution characteristics. Since $t$ represents the eddy growth step $n$ (Eq. 9), the dominant period $T_{50}$ up to which the cumulative percentage contribution to total variance is equal to 50 is obtained from Eq. (6) for value of $n$ equal to 0. In the present study of periodicities in frequency distribution of prime number at unit number spacing intervals, the primary perturbation time period $T_s$ is equal to unit number class interval and $T_{50}$ is obtained in terms of unit number class interval as

$$T_{50} = (2 + \tau)\tau^0 \approx 3.6 \qquad (28)$$

Prime numbers with spacing intervals up to 3.6 or approximately 4 contribute up to 50% to the total variance. This model prediction is in agreement with computed values of $t_{50}$ (Sec. 6.2, Fig. 10).

## 5.5. Applications of model concepts to prime number distribution

The general systems theory for fractal fluctuations in dynamical systems is based on the following concepts.

- Selfsimilar fractal fluctuations represent an eddy continuum with inherent long-range correlations manifested as $1/f$ noise where $f$ is the frequency.
- Large eddy is visualized as the envelope enclosing inherent small scale eddies. The eddy continuum is generated by successive integration of enclosed small scale eddies as given in Eq.3, namely

$$W^2 = \frac{2}{\pi} \frac{r}{R} w_*^2$$

- In *number field* domain, the above equation can be visualized as follows. The r.m.s. circulation speeds $W$ and $w_*$ are equivalent to units of computations of respective *yardstick lengths* $R$ and $r$. Spatial integration of $w_*$ units of a finite yardstick length $r$, i.e. a computational domain $w_* r$, results in a larger computational domain $WR$ (Selvam, 1993). The computed domain $WR$ is larger than the primary domain $w_* r$ because of uncertainty in the length measurement using a finite yardstick length $r$, which should be infinitesimally small in an ideal measurement. The continuum number field domain (Cartesian co-ordinates) may therefore be obtained from successive integration of enclosed finite number field domains (Selvam, 1993) as shown in Fig. 2.
- Real numbers are the computational tools used for the numerical computation of successive values of $W_n$ and $R_n$ at growth step $n$ starting from arbitrarily small computational domain $w_* r$ for length scale $r$ and $w_*$ units of computation. The successive values of $W_n$ and $R_n$ follow the Fibonacci mathematical number series. Therefore the integrated mean of the continuum number field domain at successive unit number intervals also follows the Fibonacci mathematical number series.
- Successive growth stages of the large eddy domain $W_n R_n$ (Eq. 3) traces the overall logarithmic spiral trajectory of the continuum number field domain with internal structure of the space filling quasiperiodic Penrose tiling pattern.
- Fractal structure to the frequency distribution of prime numbers at unit number intervals is associated with ordered space filling pattern of the quasiperiodic Penrose tiling pattern to the underlying continuum number field. The continuum number field therefore, may be visualized to trace a nested continuum of vortex roll circulations (whirpools of numbers) with embedded dominant eddies, the eddy lengths following Fibonacci number sequence. Positive peaks of dominant eddies signify locations of maxima for frequency of occurrence of prime numbers.

The incorporation of *Fibonacci* mathematical series, representative of ramified bifurcations, indicates ordered growth of fractal patterns (Stewart, 1992). The fractal patterns are shown to result from the cumulative integration of enclosed small-scale fluctuations (Selvam and Fadnavis, 1998; Selvam, 2007). By analogy it follows that the *continuum number field* when computed as the integrated mean over successively larger discrete domains, traces the quasiperiodic *Penrose* tiling pattern.

## 6. Data and Analysis

Lists of prime numbers in successive one million numbers for 10 million numbers were obtained from http://Prime-Numbers.org.

### 6.1 Frequency distribution of prime numbers

The frequency of occurrence of primes in *unit number class intervals*, i.e. in the form of series of 0s and 1s were determined in successive sets of 1000 numbers range each for one million numbers at a time for 10 million numbers. The first set consists of 3 to 1000 numbers excluding primes 1 and 2. Each million numbers therefore consists of 1000 sets of prime number frequency distribution. Frequency distribution of prime numbers in each 1000 numbers long data set is computed as follows.

Each 1000 number range was divided into *unit number* class intervals ranging from the first number $X_F$ to the last number $X_L$ equal to the last prime number, the number of class intervals $n$ being equal to $X_F - X_L + 1$. The domain of each class interval is *unit number* and the $n$ class intervals are $X_F$ to $X_F+1$, $X_F+1$ to $X_F+2$, etc. The midpoint of each class interval, i.e. $X_F+0.5$, $X_F+1.5$, etc., give the class interval values $x(i)$, $i=1, n$ . Any prime number X occupies a unit number domain X+1. The frequency of occurrence (0 or 1) of prime number $f(i)$ in number $n$ of class intervals $x(i)$, $i=1, n$ covers the range of values from $X_F$ [1+1000000(N-1), 1001+1000000(N-1), 2001+1000000(N-1), …….., 999001+1000000(N-1), N being the million number set 1 to 10] to $X_L$, the last prime number in the 1000 number long data set. The average *av* and standard deviation *sd* for each data set is computed as

$$av = \frac{\sum_1^n \left[ x(i) \times f(i) \right]}{\sum_1^n f(i)}$$

$$sd = \frac{\sum_1^n \left\{ [x(i) - av]^2 \times f(i) \right\}}{\sum_1^n f(i)}$$

The *normalized deviation t* values for class intervals $t(i)$ were then computed as

$$t(i) = \frac{x(i) - av}{sd}$$

The cumulative percentage probabilities of occurrence *cmax(i)* and *cmin(i)* corresponding to the *normalized deviation t* values were then computed starting respectively from the maximum ($i=n$) and minimum ($i=1$) class interval values as follows.

$$cmax(i) = \frac{\sum_{n}^{i}[x(i) \times f(i)]}{\sum_{1}^{n}[x(i) \times f(i)]} \times 100.0$$

$$cmin(i) = \frac{\sum_{1}^{i}[x(i) \times f(i)]}{\sum_{1}^{n}[x(i) \times f(i)]} \times 100.0$$

Sample calculations for cumulative percentage probability values *cmax*(*i*), *cmin*(*i*) and corresponding *normalized deviation t*(*i*) values are shown Table 3 for the first data set which starts from 3 (omitting numbers 1 and 2) and ends in 997, the last prime number within 1000 numbers. Similar computations were done for successive 1000 number range for the 1000 distributions in each million numbers. The average and standard deviation of the 1000 cumulative percentage probability values *cmax*(*i*) and *cmin*(*i*) for corresponding *normalized deviation t*(*i*) values were computed for each one million number range and plotted in Fig. 6 for the 10 million numbers. The figure also contains the statistical normal distribution and the computed theoretical distribution (Eq. 25) for comparison. The observed distributions are close to the model predicted theoretical and the statistical normal distribution for *t* values less than 2. The observed distributions are the same as the statistical normal distribution at less than or equal to 5% level of significance 'goodness of fit' as determined by the statistical *Chi-square test* (Spiegel, 1961). The power spectra of the frequency of occurrence (0 or 1) of prime number *f*(*i*) in number *n* of class intervals *x*(*i*), *i*=1, *n* also follows the universal inverse power law form of the statistical normal distribution as shown in Sec. 6.2 below and the results are consistent with model predictions derived in Sec. 4 above.

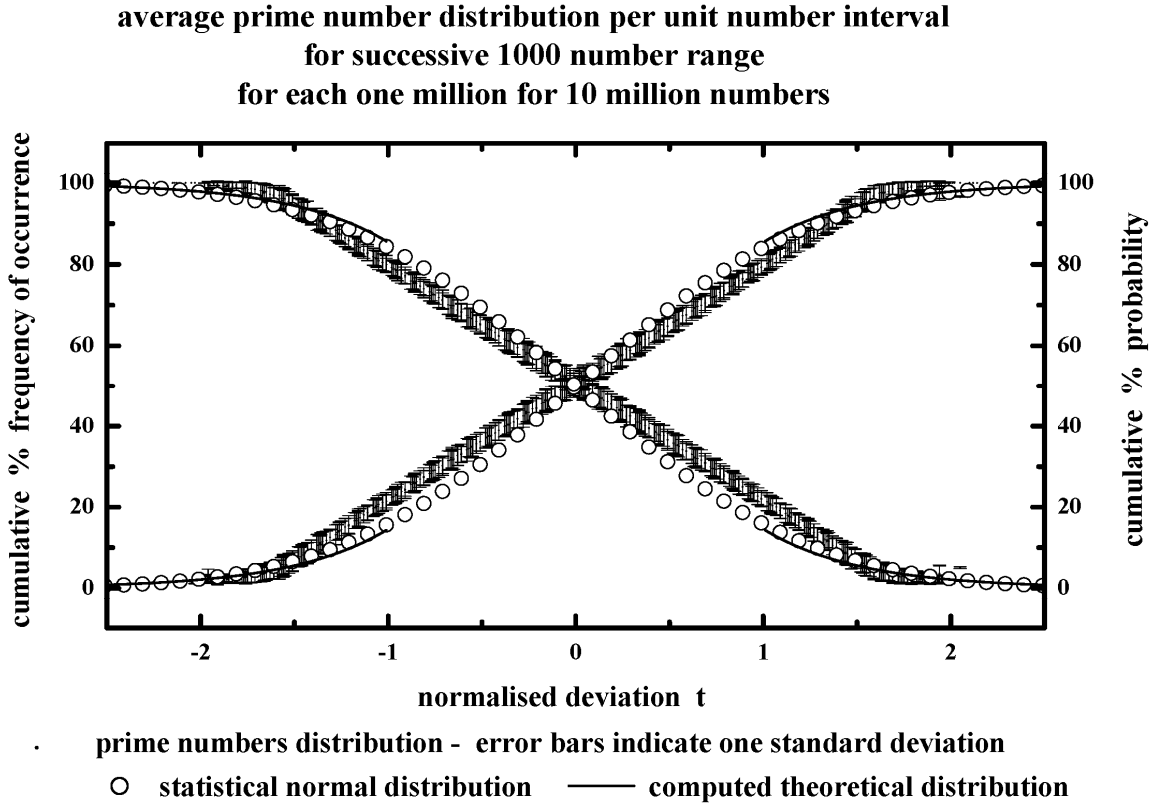

Fig. 6: Average probability density distribution of prime numbers in unit number class intervals for successive 1000 numbers for each one million numbers (1000 sets) for 10 million numbers. Error bars indicate one standard deviation on either side of the mean. Statistical normal distribution (open circles) and computed theoretical distribution (line) are also shown in the figure. The computed theoretical distribution is close to the statistical normal distribution for normalized deviation $t$ values less than 2 in the range of observed values for the distribution.

## 6.2 Power Spectral Analysis: Analyses Techniques, Data and Results

The broadband power spectrum of space-time fluctuations of dynamical systems can be computed accurately by an elementary, but very powerful method of analysis developed by Jenkinson (1977) which provides a quasi-continuous form of the classical periodogram allowing systematic allocation of the total variance and degrees of freedom of the data series to logarithmically spaced elements of the frequency range (0.5, 0). The periodogram is constructed for a fixed set of $10000(m)$ periodicities $L_m$ which increase geometrically as $L_m=2$ $\exp(Cm)$ where $C=.001$ and $m=0, 1, 2,.....m$ . The data series $Y_t$ for the $N$ data points was used. The periodogram estimates the set of $A_m\cos(2\pi v_m S\text{-}\phi_m)$ where $A_m$, $v_m$ and $\phi_m$ denote respectively the amplitude, frequency and phase angle for the $m^{th}$ periodicity and $S$ is the time or space interval. The data used in the present study is the same as that described in Section 6.1, namely, the frequency ($f$ ($i$), $i=1$, $n$) of occurrence of primes in *unit number class intervals* in successive 1000 number range, generating 1000 data sets for each million numbers for a total of 10 million numbers. The cumulative percentage contribution to total variance was computed starting from the high frequency side of the spectrum. The period $T_{50}$ at which 50% contribution to total variance occurs is taken as reference and the normalized standard deviation $t_m$ values are computed as (Eq. 27).

$$t_m = \left( \frac{\log L_m}{\log T_{50}} \right) - 1$$

The cumulative percentage contribution to total variance, the cumulative percentage normalized phase (normalized with respect to the total phase rotation) and the corresponding $t$ values were computed. The power spectra were plotted as cumulative percentage contribution to total variance versus the normalized standard deviation $t$ as given above. The period $L$ is in class interval units which is equal to *one* (unit number) in the present study. Periodicities up to $T_{50}$ contribute up to 50% of total variance. The phase spectra were plotted as cumulative (%) normalized (normalized to total rotation) phase .The average and standard deviation for the 1000 sets of variance and phase spectra were computed for each one million numbers and plotted in Fig. 7 for the 10 million numbers used in the study along with statistical normal distribution and computed theoretical distribution. The computed theoretical distribution is very close to the statistical normal distribution for normalized standard deviation $t$ less than 2, covering the range of observed values for the power spectra. The 'goodness of fit' (Spiegel, 1961) between the variance spectrum, phase spectrum and statistical normal distribution is significant at <= 5% level for all the sets.

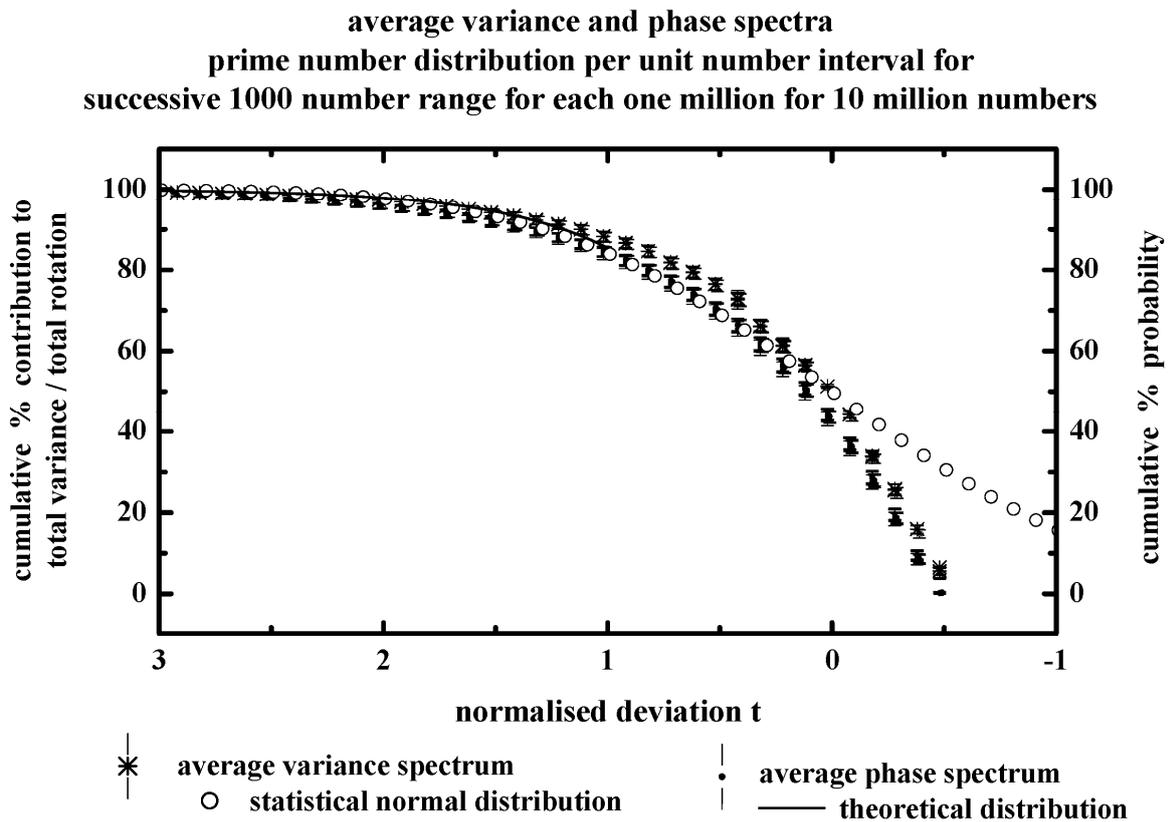

Fig. 7: The average variance and phase spectra. Standard deviations are shown as vertical error bars. Statistical normal distribution (open circles) and computed theoretical distribution (line) are also shown.

The peak wavelength (period) for dominant significant (less than 5%) wavebands and the corresponding percentage contribution to total variance by the waveband for all the data sets are shown in Figs. 8 and 9 respectively for wavelengths (periodicities) 2 to 10 and 10 to 30. It is seen that periodicities 2, 3 and 6 are present in all the data sets and contribute at a markedly higher level (2 to 7 %) to the total variance

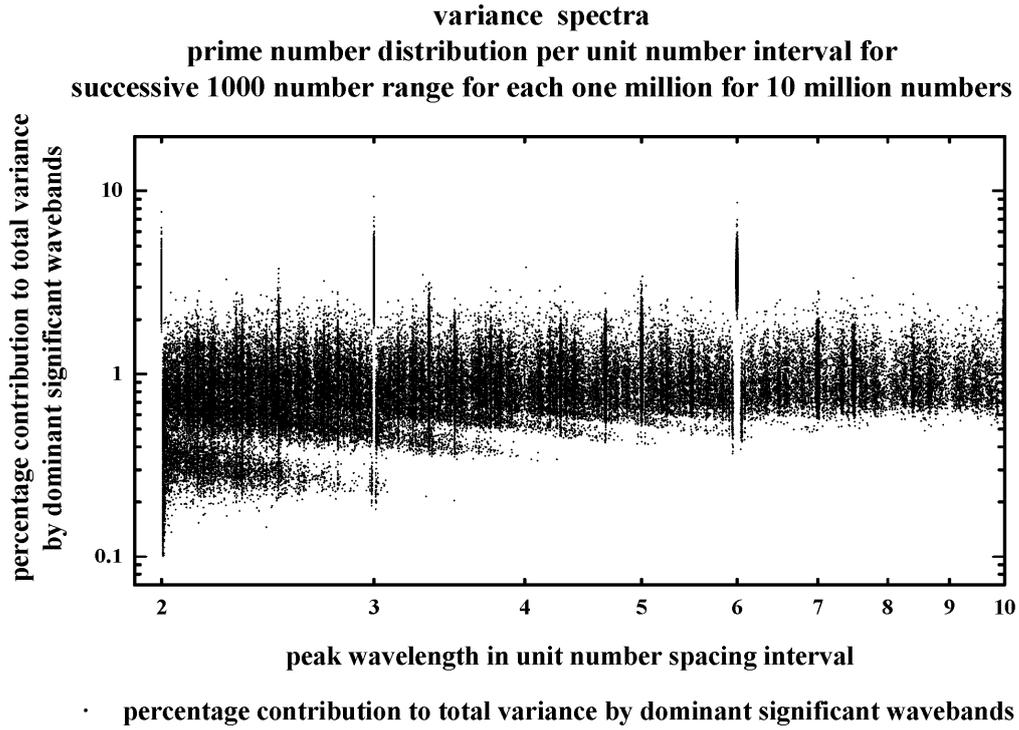

Fig. 8: Dominant (normalised variance greater than 1) statistically significant wavebands. Peak periodicities and percentage contribution to total variance by the wavebands

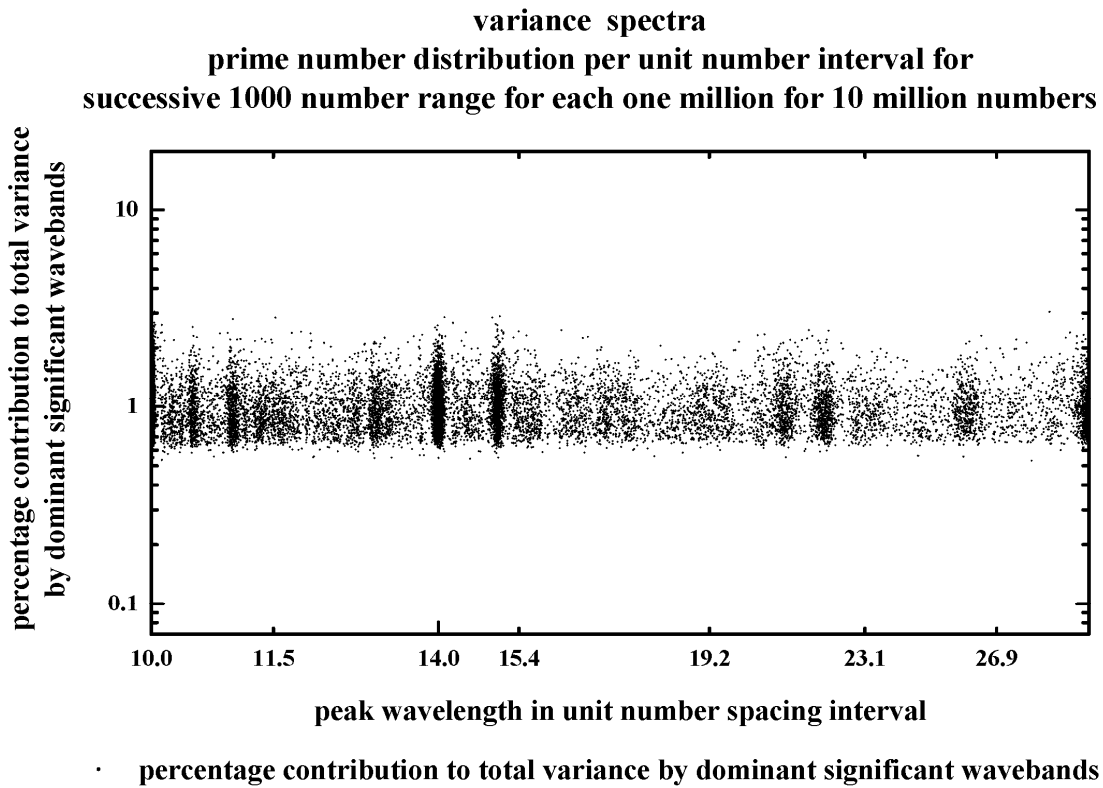

Fig. 9: Dominant (normalised variance greater than 1) statistically significant wavebands. Peak periodicities and percentage contribution to total variance by the wavebands

The dominant peak periodicities are in close agreement with model predicted dominant peak periodicities, e.g. *2.2, 3.6, 5.8, 9.5, 15.3, 24.8, 40.1,* and *64.9* prime number spacing intervals for values of *n* ranging from -1 to 6 (Eq. 6). The normalised variance and phase spectra follow each other closely (the '*goodness of fit* ' being significant at <= 5%) displaying *Berry* 's phase in the quantum-like chaos exhibited by prime number distribution. Earlier study by Marek Wolf (May 1996, IFTUWr 908/96 http://rose.ift.uni.wroc.pl/~mwolf) also shows that the number of *Twins* (spacing interval 2) and primes separated by a gap of length 4 ("*cousins*") is almost the same and it determines a fractal structure on the set of primes. The conjecture that there should be approximately equal numbers of prime power pairs differing by 2 and by 4, but about twice as many differing by 6 is proved to be true by Gopalkrishna Gadiyar and Padma (1999 http://www.maths.ex.ac.uk/~mwatkins/zeta/padma.pdf). The dominant perodicities shown above at Figs. 8 and 9 are consistent with these reported results.

Dahmen *et al* (2001) present numerical evidence for regularities in the distribution of gaps between primes when these are divided into congruence families (in Dirichlet's classification). The histograms for the distribution of gaps of families are scale invariant. The number of gaps of size d=2 and 4 is asymptotically equal. This was first conjectured by Hardy and Littlewood (1923) in a seminal paper almost 80 years ago and numerically verified by Wolf (1999 and references therein) for a large number of primes. The calculations of Dahmen *et al* (2001) however, show that these results hold for gaps of size d = 4 and 8, d = 8 and 16, d = 16 and 32. Also, there are peaks at gaps of size d= multiple of 6 (a result observed by Wolf in the 2-family). Dahmen *et al*'s results indicate that this property extends to other k-families as long as k is a power of 2. There is still no explanation why these regularities appear.

Kumar *et al* (2008) studied the statistical properties of the distances (difference between two consecutive prime numbers) and their increments (the difference between two consecutive distances) for a sequence comprising the first $5 \times 10^7$ prime numbers. They find that the histogram of the increments follows an exponential distribution with superposed periodic behavior of period three, similar to previously-reported (Wolf, 1999) period six oscillations for the distances between consecutive primes.

Scafetta et al (2004) studied the fractal properties of the distances between consecutive primes. The distance sequence is found to be well described by a non-stationary exponential probability distribution. They propose an intensity-expansion method to treat this non-stationarity and find that the statistics underlying the distance between consecutive primes is Gaussian and that, by transforming the distance sequence into a stationary one, the range of Gaussian randomness of the sequence increases.

**prime number frequency distribution per unit number interval**
**$T_{50}$ values for successive 1000 number range**
**for each one million for 10 million numbers**

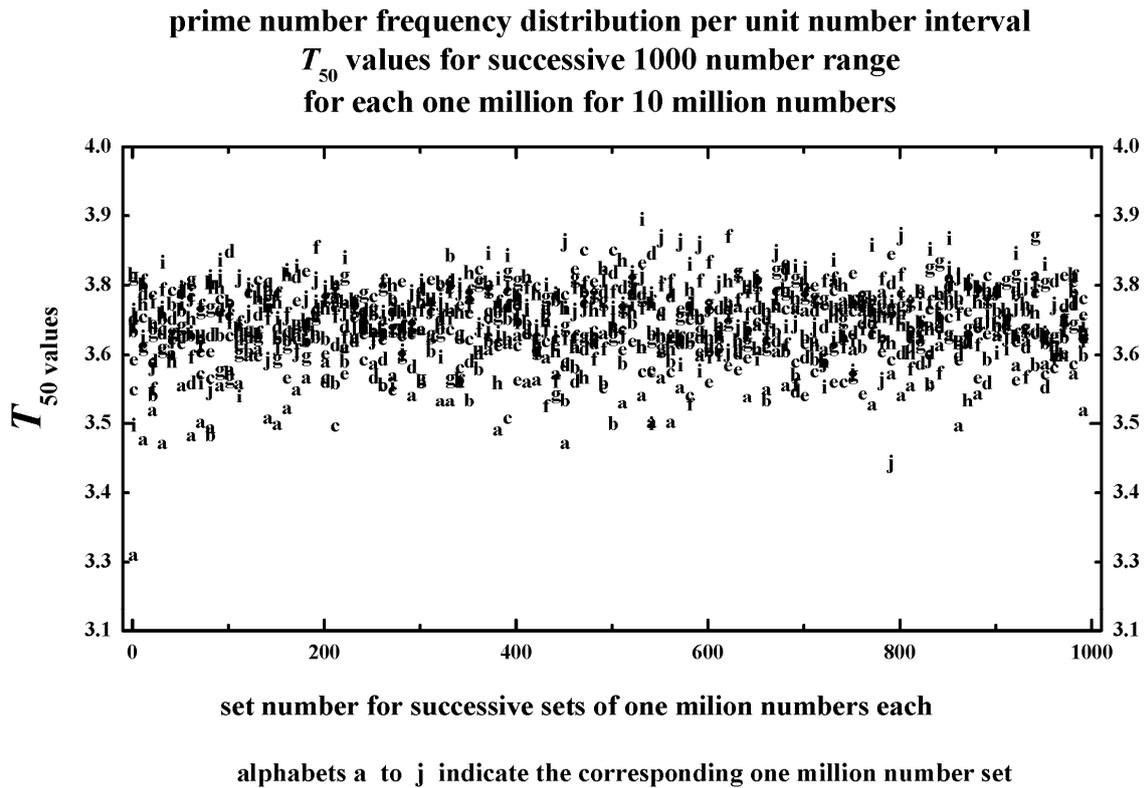

**set number for successive sets of one milion numbers each**

**alphabets a to j indicate the corresponding one million number set**

Fig. 10 Periodicities $t_{50}$ values upto which the cumulative percentage contribution to total variance is equal to 50 for the successive 1000 number sets in each one million numbers for 10 million numbers; the alphabets a to j refer to the 1 to 10 million number sets

The period $T_{50}$ up to which the cumulative percentage contribution to total variance is equal to 50 for all the data sets is plotted in Fig. 10 and the values are in approximate agreement with model predicted value of $T_{50}$ equal to about *3.6* (Eq. 28). The dominant significant period 2 corresponds to *twin primes*. In *number theory* [Rose, 1995; Beiler, 1966] the *twin prime conjecture* states that there are many pairs of primes *p, q* where *q = p + 2*. There are infinitely many prime pairs as *z* tends to infinity.

The average and standard deviation of the frequency ($f(i)$, $i=1$, $n$), described in Sec. 6.1 of prime number distribution at unit numbers intervals used for the spectral analyses are plotted in Fig. 11 for the 1000 sets in each one million numbers for the 10 million numbers used for the study. The average and standard deviations for the different number ranges are more or less the same.

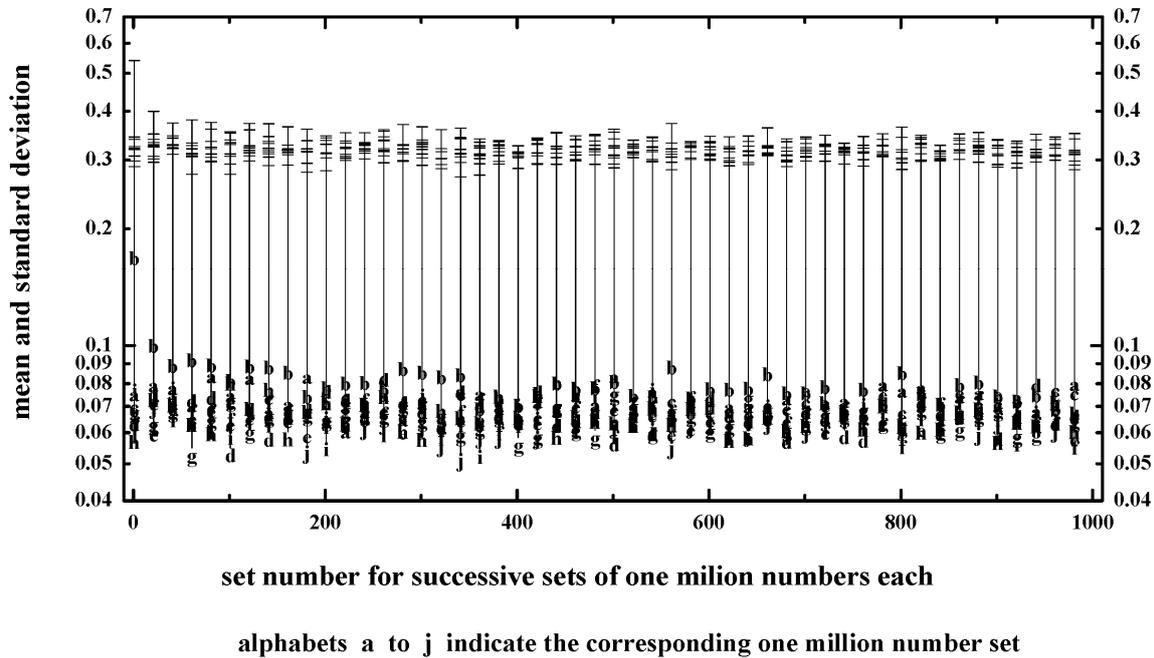

**prime number frequency distribution per unit number interval average and standard deviation for successive 1000 number range for each one million for 10 million numbers**

**set number for successive sets of one milion numbers each**

**alphabets a to j indicate the corresponding one million number set**

Fig. 11 The mean and standard deviation of frequency distribution of prime numbers at unit number class intervals for each 1000 numbers data set. The number of data sets is 1000 for each one million numbers. The alphabets 'a' to 'j' indicate the successive one million for the 10 million numbers

## 7. Discussion and conclusions

In mathematics, Cantorian fractal space-time fluctuations is now associated with reference to quantum systems (Ord, 1983; Nottale, 1989; El Naschie, 1993, 1998, 2008). Recent studies indicate a close association between *number theory* in mathematics, in particular, the distribution of *prime numbers* and the chaotic orbits of excited quantum systems such as the hydrogen atom (Cipra, 1996; Berry, 1992; Cipra http://www.maths.ex.ac.uk/~mwatkins/zeta/cipra.htm). The spacing intervals of adjacent prime numbers exhibit *fractal* fluctuations generic to diverse dynamical systems in nature. The irregular (chaotic) *fractal* fluctuations however, exhibit *self-similar* geometry manifested as inverse power-law form for power spectra. *Self-similar* fluctuations imply long-range correlations or non-local connections identified as *self-organized criticality*. A general systems theory model for atmospheric flows developed by the author shows that *self-organized criticality* is a signature of quantum-like chaos. The general systems theory is applicable to all dynamical systems in nature. An earlier (Selvam, 2001) study applying the model concepts show that quantum-like chaos in dynamical systems incorporates *prime number distribution functions* in the quantification of *self-organized criticality*.

The model predictions are as follows: (i) The probability distribution and power spectrum of fractals follow the same inverse power law which is a function of the *golden mean*. The predicted inverse power law distribution is very close to the statistical normal distribution for fluctuations within two standard deviations from the mean of the distribution. (ii) Fractals signify quantumlike chaos since variance spectrum represents probability density distribution, a characteristic of quantum systems such as electron or photon. (ii) Fractal fluctuations of frequency distribution of prime numbers signify spontaneous organisation of underlying continuum number field into the ordered pattern of the quasiperiodic Penrose tiling pattern. The model predictions are in agreement with the probability distributions and power spectra for different sets of frequency of occurrence of prime numbers at unit number interval for successive 1000 numbers. Prime numbers in the first 10 million numbers were used for the study.

## Acknowledgements

The author is grateful to Dr. A. S. R. Murty for his keen interest and encouragement during the course of the study.

# Appendix I

| Table 3 Unit number class interval-wise frequency distribution of prime numbers within 3 to 997 (last prime number within 1000 numbers) | | | | | | | | | |
|---|---|---|---|---|---|---|---|---|---|
| no | class | freq | cmin | cmax | t | cmin% | normal | cmax% | normal |
| 1 | 3.50 | 1 | 1 | 167 | 1.53 | .60 | 6.30 | 100.00 | 93.70 |
| 2 | 4.50 | 1 | 2 | 166 | 1.52 | 1.20 | 6.43 | 99.40 | 93.57 |
| 4 | 6.50 | 1 | 3 | 165 | 1.52 | 1.80 | 6.43 | 98.80 | 93.57 |
| 8 | 10.50 | 1 | 4 | 164 | 1.50 | 2.39 | 6.68 | 98.20 | 93.32 |
| 10 | 12.50 | 1 | 5 | 163 | -1.50 | 2.99 | 6.68 | 97.61 | 93.32 |
| 14 | 16.50 | 1 | 6 | 162 | -1.48 | 3.59 | 6.94 | 97.01 | 93.06 |
| 16 | 18.50 | 1 | 7 | 161 | -1.48 | 4.19 | 6.94 | 96.41 | 93.06 |
| 20 | 22.50 | 1 | 8 | 160 | -1.46 | 4.79 | 7.22 | 95.81 | 92.79 |
| 26 | 28.50 | 1 | 9 | 159 | -1.44 | 5.39 | 7.49 | 95.21 | 92.51 |
| 28 | 30.50 | 1 | 10 | 158 | -1.43 | 5.99 | 7.64 | 94.61 | 92.36 |
| 34 | 36.50 | 1 | 11 | 157 | -1.41 | 6.59 | 7.93 | 94.01 | 92.07 |
| 38 | 40.50 | 1 | 12 | 156 | -1.40 | 7.19 | 8.08 | 93.41 | 91.92 |
| 40 | 42.50 | 1 | 13 | 155 | -1.39 | 7.78 | 8.23 | 92.81 | 91.77 |
| 44 | 46.50 | 1 | 14 | 154 | -1.38 | 8.38 | 8.38 | 92.22 | 91.62 |
| 50 | 52.50 | 1 | 15 | 153 | -1.36 | 8.98 | 8.69 | 91.62 | 91.31 |
| 56 | 58.50 | 1 | 16 | 152 | -1.34 | 9.58 | 9.01 | 91.02 | 90.99 |
| 58 | 60.50 | 1 | 17 | 151 | -1.33 | 10.18 | 9.18 | 90.42 | 90.82 |
| 64 | 66.50 | 1 | 18 | 150 | -1.31 | 10.78 | 9.51 | 89.82 | 90.49 |
| 68 | 70.50 | 1 | 19 | 149 | -1.30 | 11.38 | 9.68 | 89.22 | 90.32 |
| 70 | 72.50 | 1 | 20 | 148 | -1.29 | 11.98 | 9.85 | 88.62 | 90.15 |
| 76 | 78.50 | 1 | 21 | 147 | -1.27 | 12.57 | 10.20 | 88.02 | 89.80 |
| 80 | 82.50 | 1 | 22 | 146 | -1.26 | 13.17 | 10.38 | 87.43 | 89.62 |
| 86 | 88.50 | 1 | 23 | 145 | -1.24 | 13.77 | 10.75 | 86.83 | 89.25 |
| 94 | 96.50 | 1 | 24 | 144 | -1.21 | 14.37 | 11.31 | 86.23 | 88.69 |
| 98 | 100.50 | 1 | 25 | 143 | -1.20 | 14.97 | 11.51 | 85.63 | 88.49 |
| 100 | 102.50 | 1 | 26 | 142 | -1.19 | 15.57 | 11.70 | 85.03 | 88.30 |
| 104 | 106.50 | 1 | 27 | 141 | -1.18 | 16.17 | 11.90 | 84.43 | 88.10 |
| 106 | 108.50 | 1 | 28 | 140 | -1.17 | 16.77 | 12.10 | 83.83 | 87.90 |
| 110 | 112.50 | 1 | 29 | 139 | -1.16 | 17.36 | 12.30 | 83.23 | 87.70 |
| 124 | 126.50 | 1 | 30 | 138 | -1.11 | 17.96 | 13.35 | 82.64 | 86.65 |
| 128 | 130.50 | 1 | 31 | 137 | -1.10 | 18.56 | 13.57 | 82.04 | 86.43 |
| 134 | 136.50 | 1 | 32 | 136 | -1.08 | 19.16 | 14.01 | 81.44 | 85.99 |
| 136 | 138.50 | 1 | 33 | 135 | -1.07 | 19.76 | 14.23 | 80.84 | 85.77 |
| 146 | 148.50 | 1 | 34 | 134 | -1.04 | 20.36 | 14.92 | 80.24 | 85.08 |
| 148 | 150.50 | 1 | 35 | 133 | -1.03 | 20.96 | 15.15 | 79.64 | 84.85 |
| 154 | 156.50 | 1 | 36 | 132 | -1.01 | 21.56 | 15.63 | 79.04 | 84.38 |
| 160 | 162.50 | 1 | 37 | 131 | -.99 | 22.16 | 16.11 | 78.44 | 83.89 |
| 164 | 166.50 | 1 | 38 | 130 | -.98 | 22.75 | 16.35 | 77.84 | 83.65 |
| 170 | 172.50 | 1 | 39 | 129 | -.96 | 23.35 | 16.85 | 77.25 | 83.15 |
| 176 | 178.50 | 1 | 40 | 128 | -.93 | 23.95 | 17.62 | 76.65 | 82.38 |
| 178 | 180.50 | 1 | 41 | 127 | -.93 | 24.55 | 17.62 | 76.05 | 82.38 |
| 188 | 190.50 | 1 | 42 | 126 | -.89 | 25.15 | 18.67 | 75.45 | 81.33 |
| 190 | 192.50 | 1 | 43 | 125 | -.89 | 25.75 | 18.67 | 74.85 | 81.33 |
| 194 | 196.50 | 1 | 44 | 124 | -.87 | 26.35 | 19.22 | 74.25 | 80.79 |
| 196 | 198.50 | 1 | 45 | 123 | -.87 | 26.95 | 19.22 | 73.65 | 80.79 |
| 208 | 210.50 | 1 | 46 | 122 | -.83 | 27.55 | 20.33 | 73.05 | 79.67 |
| 220 | 222.50 | 1 | 47 | 121 | -.79 | 28.14 | 21.48 | 72.46 | 78.52 |
| 224 | 226.50 | 1 | 48 | 120 | -.77 | 28.74 | 22.07 | 71.86 | 77.93 |
| 226 | 228.50 | 1 | 49 | 119 | -.77 | 29.34 | 22.07 | 71.26 | 77.93 |
| 230 | 232.50 | 1 | 50 | 118 | -.75 | 29.94 | 22.66 | 70.66 | 77.34 |
| 236 | 238.50 | 1 | 51 | 117 | -.73 | 30.54 | 23.27 | 70.06 | 76.73 |
| 238 | 240.50 | 1 | 52 | 116 | -.73 | 31.14 | 23.27 | 69.46 | 76.73 |
| 248 | 250.50 | 1 | 53 | 115 | -.69 | 31.74 | 24.51 | 68.86 | 75.49 |
| 254 | 256.50 | 1 | 54 | 114 | -.67 | 32.33 | 25.14 | 68.26 | 74.86 |
| 260 | 262.50 | 1 | 55 | 113 | -.65 | 32.93 | 25.78 | 67.67 | 74.21 |
| 266 | 268.50 | 1 | 56 | 112 | -.63 | 33.53 | 26.43 | 67.07 | 73.57 |
| 268 | 270.50 | 1 | 57 | 111 | -.62 | 34.13 | 26.76 | 66.47 | 73.24 |

Table 3 Unit number class interval-wise frequency distribution of prime numbers within 3 to 997 (last prime number within 1000 numbers)

| no | class | freq | cmin | cmax | t | cmin% | normal | cmax% | normal |
|---|---|---|---|---|---|---|---|---|---|
| 274 | 276.50 | 1 | 58 | 110 | -.60 | 34.73 | 27.42 | 65.87 | 72.57 |
| 278 | 280.50 | 1 | 59 | 109 | -.59 | 35.33 | 27.76 | 65.27 | 72.24 |
| 280 | 282.50 | 1 | 60 | 108 | -.58 | 35.93 | 28.10 | 64.67 | 71.90 |
| 290 | 292.50 | 1 | 61 | 107 | -.55 | 36.53 | 29.12 | 64.07 | 70.88 |
| 304 | 306.50 | 1 | 62 | 106 | -.50 | 37.13 | 30.85 | 63.47 | 69.15 |
| 308 | 310.50 | 1 | 63 | 105 | -.49 | 37.72 | 31.21 | 62.87 | 68.79 |
| 310 | 312.50 | 1 | 64 | 104 | -.48 | 38.32 | 31.56 | 62.28 | 68.44 |
| 314 | 316.50 | 1 | 65 | 103 | -.47 | 38.92 | 31.92 | 61.68 | 68.08 |
| 328 | 330.50 | 1 | 66 | 102 | -.42 | 39.52 | 33.72 | 61.08 | 66.28 |
| 334 | 336.50 | 1 | 67 | 101 | -.40 | 40.12 | 34.46 | 60.48 | 65.54 |
| 344 | 346.50 | 1 | 68 | 100 | -.37 | 40.72 | 35.57 | 59.88 | 64.43 |
| 346 | 348.50 | 1 | 69 | 99 | -.36 | 41.32 | 35.94 | 59.28 | 64.06 |
| 350 | 352.50 | 1 | 70 | 98 | -.35 | 41.92 | 36.32 | 58.68 | 63.68 |
| 356 | 358.50 | 1 | 71 | 97 | -.33 | 42.51 | 37.07 | 58.08 | 62.93 |
| 364 | 366.50 | 1 | 72 | 96 | -.30 | 43.11 | 38.21 | 57.49 | 61.79 |
| 370 | 372.50 | 1 | 73 | 95 | -.28 | 43.71 | 38.97 | 56.89 | 61.03 |
| 376 | 378.50 | 1 | 74 | 94 | -.26 | 44.31 | 39.74 | 56.29 | 60.26 |
| 380 | 382.50 | 1 | 75 | 93 | -.25 | 44.91 | 40.13 | 55.69 | 59.87 |
| 386 | 388.50 | 1 | 76 | 92 | -.23 | 45.51 | 40.90 | 55.09 | 59.10 |
| 394 | 396.50 | 1 | 77 | 91 | -.20 | 46.11 | 42.07 | 54.49 | 57.93 |
| 398 | 400.50 | 1 | 78 | 90 | -.19 | 46.71 | 42.47 | 53.89 | 57.53 |
| 406 | 408.50 | 1 | 79 | 89 | -.16 | 47.31 | 43.64 | 53.29 | 56.36 |
| 416 | 418.50 | 1 | 80 | 88 | -.12 | 47.90 | 45.22 | 52.69 | 54.78 |
| 418 | 420.50 | 1 | 81 | 87 | -.12 | 48.50 | 45.22 | 52.10 | 54.78 |
| 428 | 430.50 | 1 | 82 | 86 | -.08 | 49.10 | 46.81 | 51.50 | 53.19 |
| 430 | 432.50 | 1 | 83 | 85 | -.08 | 49.70 | 46.81 | 50.90 | 53.19 |
| 436 | 438.50 | 1 | 84 | 84 | -.06 | 50.30 | 47.61 | 50.30 | 52.39 |
| 440 | 442.50 | 1 | 85 | 83 | -.04 | 50.90 | 48.40 | 49.70 | 51.60 |
| 446 | 448.50 | 1 | 86 | 82 | -.02 | 51.50 | 49.20 | 49.10 | 50.80 |
| 454 | 456.50 | 1 | 87 | 81 | .00 | 52.10 | 50.00 | 48.50 | 50.00 |
| 458 | 460.50 | 1 | 88 | 80 | .02 | 52.69 | 50.80 | 47.90 | 49.20 |
| 460 | 462.50 | 1 | 89 | 79 | .02 | 53.29 | 50.80 | 47.31 | 49.20 |
| 464 | 466.50 | 1 | 90 | 78 | .04 | 53.89 | 51.60 | 46.71 | 48.40 |
| 476 | 478.50 | 1 | 91 | 77 | .08 | 54.49 | 53.19 | 46.11 | 46.81 |
| 484 | 486.50 | 1 | 92 | 76 | .11 | 55.09 | 54.38 | 45.51 | 45.62 |
| 488 | 490.50 | 1 | 93 | 75 | .12 | 55.69 | 54.78 | 44.91 | 45.22 |
| 496 | 498.50 | 1 | 94 | 74 | .15 | 56.29 | 55.96 | 44.31 | 44.04 |
| 500 | 502.50 | 1 | 95 | 73 | .16 | 56.89 | 56.36 | 43.71 | 43.64 |
| 506 | 508.50 | 1 | 96 | 72 | .18 | 57.49 | 57.14 | 43.11 | 42.86 |
| 518 | 520.50 | 1 | 97 | 71 | .22 | 58.08 | 58.71 | 42.51 | 41.29 |
| 520 | 522.50 | 1 | 98 | 70 | .23 | 58.68 | 59.10 | 41.92 | 40.90 |
| 538 | 540.50 | 1 | 99 | 69 | .29 | 59.28 | 61.41 | 41.32 | 38.59 |
| 544 | 546.50 | 1 | 100 | 68 | .31 | 59.88 | 62.17 | 40.72 | 37.83 |
| 554 | 556.50 | 1 | 101 | 67 | .34 | 60.48 | 63.31 | 40.12 | 36.69 |
| 560 | 562.50 | 1 | 102 | 66 | .36 | 61.08 | 64.06 | 39.52 | 35.94 |
| 566 | 568.50 | 1 | 103 | 65 | .38 | 61.68 | 64.80 | 38.92 | 35.20 |
| 568 | 570.50 | 1 | 104 | 64 | .39 | 62.28 | 65.17 | 38.32 | 34.83 |
| 574 | 576.50 | 1 | 105 | 63 | .41 | 62.87 | 65.91 | 37.72 | 34.09 |
| 584 | 586.50 | 1 | 106 | 62 | .44 | 63.47 | 67.00 | 37.13 | 33.00 |
| 590 | 592.50 | 1 | 107 | 61 | .46 | 64.07 | 67.72 | 36.53 | 32.28 |
| 596 | 598.50 | 1 | 108 | 60 | .48 | 64.67 | 68.44 | 35.93 | 31.56 |
| 598 | 600.50 | 1 | 109 | 59 | .49 | 65.27 | 68.79 | 35.33 | 31.21 |
| 604 | 606.50 | 1 | 110 | 58 | .51 | 65.87 | 69.50 | 34.73 | 30.50 |
| 610 | 612.50 | 1 | 111 | 57 | .53 | 66.47 | 70.19 | 34.13 | 29.81 |
| 614 | 616.50 | 1 | 112 | 56 | .54 | 67.07 | 70.54 | 33.53 | 29.46 |
| 616 | 618.50 | 1 | 113 | 55 | .55 | 67.67 | 70.88 | 32.93 | 29.12 |
| 628 | 630.50 | 1 | 114 | 54 | .59 | 68.26 | 72.24 | 32.33 | 27.76 |
| 638 | 640.50 | 1 | 115 | 53 | .63 | 68.86 | 73.57 | 31.74 | 26.43 |
| 640 | 642.50 | 1 | 116 | 52 | .63 | 69.46 | 73.57 | 31.14 | 26.43 |
| 644 | 646.50 | 1 | 117 | 51 | .65 | 70.06 | 74.21 | 30.54 | 25.78 |

Table 3 Unit number class interval-wise frequency distribution of prime numbers within 3 to 997 (last prime number within 1000 numbers)

| no | class | freq | cmin | cmax | t | cmin% | normal | cmax% | normal |
|---|---|---|---|---|---|---|---|---|---|
| 650 | 652.50 | 1 | 118 | 50 | .67 | 70.66 | 74.86 | 29.94 | 25.14 |
| 656 | 658.50 | 1 | 119 | 49 | .69 | 71.26 | 75.49 | 29.34 | 24.51 |
| 658 | 660.50 | 1 | 120 | 48 | .69 | 71.86 | 75.49 | 28.74 | 24.51 |
| 670 | 672.50 | 1 | 121 | 47 | .73 | 72.46 | 76.73 | 28.14 | 23.27 |
| 674 | 676.50 | 1 | 122 | 46 | .75 | 73.05 | 77.34 | 27.55 | 22.66 |
| 680 | 682.50 | 1 | 123 | 45 | .77 | 73.65 | 77.93 | 26.95 | 22.07 |
| 688 | 690.50 | 1 | 124 | 44 | .79 | 74.25 | 78.52 | 26.35 | 21.48 |
| 698 | 700.50 | 1 | 125 | 43 | .83 | 74.85 | 79.67 | 25.75 | 20.33 |
| 706 | 708.50 | 1 | 126 | 42 | .85 | 75.45 | 80.23 | 25.15 | 19.77 |
| 716 | 718.50 | 1 | 127 | 41 | .89 | 76.05 | 81.33 | 24.55 | 18.67 |
| 724 | 726.50 | 1 | 128 | 40 | .92 | 76.65 | 82.12 | 23.95 | 17.88 |
| 730 | 732.50 | 1 | 129 | 39 | .94 | 77.25 | 82.64 | 23.35 | 17.36 |
| 736 | 738.50 | 1 | 130 | 38 | .96 | 77.84 | 83.15 | 22.75 | 16.85 |
| 740 | 742.50 | 1 | 131 | 37 | .97 | 78.44 | 83.40 | 22.16 | 16.60 |
| 748 | 750.50 | 1 | 132 | 36 | 1.00 | 79.04 | 84.14 | 21.56 | 15.87 |
| 754 | 756.50 | 1 | 133 | 35 | 1.02 | 79.64 | 84.61 | 20.96 | 15.39 |
| 758 | 760.50 | 1 | 134 | 34 | 1.03 | 80.24 | 84.85 | 20.36 | 15.15 |
| 766 | 768.50 | 1 | 135 | 33 | 1.06 | 80.84 | 85.54 | 19.76 | 14.46 |
| 770 | 772.50 | 1 | 136 | 32 | 1.07 | 81.44 | 85.77 | 19.16 | 14.23 |
| 784 | 786.50 | 1 | 137 | 31 | 1.12 | 82.04 | 86.86 | 18.56 | 13.14 |
| 794 | 796.50 | 1 | 138 | 30 | 1.15 | 82.64 | 87.49 | 17.96 | 12.51 |
| 806 | 808.50 | 1 | 139 | 29 | 1.19 | 83.23 | 88.30 | 17.36 | 11.70 |
| 808 | 810.50 | 1 | 140 | 28 | 1.20 | 83.83 | 88.49 | 16.77 | 11.51 |
| 818 | 820.50 | 1 | 141 | 27 | 1.23 | 84.43 | 89.07 | 16.17 | 10.94 |
| 820 | 822.50 | 1 | 142 | 26 | 1.24 | 85.03 | 89.25 | 15.57 | 10.75 |
| 824 | 826.50 | 1 | 143 | 25 | 1.25 | 85.63 | 89.43 | 14.97 | 10.56 |
| 826 | 828.50 | 1 | 144 | 24 | 1.26 | 86.23 | 89.62 | 14.37 | 10.38 |
| 836 | 838.50 | 1 | 145 | 23 | 1.29 | 86.83 | 90.15 | 13.77 | 9.85 |
| 850 | 852.50 | 1 | 146 | 22 | 1.34 | 87.43 | 90.99 | 13.17 | 9.01 |
| 854 | 856.50 | 1 | 147 | 21 | 1.35 | 88.02 | 91.15 | 12.57 | 8.85 |
| 856 | 858.50 | 1 | 148 | 20 | 1.36 | 88.62 | 91.31 | 11.98 | 8.69 |
| 860 | 862.50 | 1 | 149 | 19 | 1.37 | 89.22 | 91.47 | 11.38 | 8.53 |
| 874 | 876.50 | 1 | 150 | 18 | 1.42 | 89.82 | 92.22 | 10.78 | 7.78 |
| 878 | 880.50 | 1 | 151 | 17 | 1.44 | 90.42 | 92.51 | 10.18 | 7.49 |
| 880 | 882.50 | 1 | 152 | 16 | 1.44 | 91.02 | 92.51 | 9.58 | 7.49 |
| 884 | 886.50 | 1 | 153 | 15 | 1.46 | 91.62 | 92.79 | 8.98 | 7.22 |
| 904 | 906.50 | 1 | 154 | 14 | 1.52 | 92.22 | 93.57 | 8.38 | 6.43 |
| 908 | 910.50 | 1 | 155 | 13 | 1.54 | 92.81 | 93.82 | 7.78 | 6.18 |
| 916 | 918.50 | 1 | 156 | 12 | 1.56 | 93.41 | 94.06 | 7.19 | 5.94 |
| 926 | 928.50 | 1 | 157 | 11 | 1.60 | 94.01 | 94.52 | 6.59 | 5.48 |
| 934 | 936.50 | 1 | 158 | 10 | 1.62 | 94.61 | 94.74 | 5.99 | 5.26 |
| 938 | 940.50 | 1 | 159 | 9 | 1.64 | 95.21 | 94.95 | 5.39 | 5.05 |
| 944 | 946.50 | 1 | 160 | 8 | 1.66 | 95.81 | 95.15 | 4.79 | 4.85 |
| 950 | 952.50 | 1 | 161 | 7 | 1.68 | 96.41 | 95.35 | 4.19 | 4.65 |
| 964 | 966.50 | 1 | 162 | 6 | 1.73 | 97.01 | 95.82 | 3.59 | 4.18 |
| 968 | 970.50 | 1 | 163 | 5 | 1.74 | 97.61 | 95.91 | 2.99 | 4.09 |
| 974 | 976.50 | 1 | 164 | 4 | 1.76 | 98.20 | 96.08 | 2.39 | 3.92 |
| 980 | 982.50 | 1 | 165 | 3 | 1.78 | 98.80 | 96.25 | 1.80 | 3.75 |
| 988 | 990.50 | 1 | 166 | 2 | 1.81 | 99.40 | 96.49 | 1.20 | 3.52 |
| 994 | 996.50 | 1 | 167 | 1 | 1.83 | 100.00 | 96.64 | .60 | 3.36 |

Distribution average ($av$) = 455.3; standard deviation ($sd$) = 296.12; number of class intervals ($n$) = 994